\documentclass[manuscript]{acmart}

\usepackage{newtxmath}
\usepackage{graphicx}
\graphicspath{{figures/}{./}}
\usepackage{amsfonts}
\usepackage{booktabs}
\usepackage{multirow}
\usepackage{microtype}
\usepackage{xcolor}
\usepackage[most]{tcolorbox}
\usepackage{pifont}
\usepackage{tikz}
\usepackage{siunitx}
\usetikzlibrary{arrows.meta,positioning,calc,fit,backgrounds,shapes.geometric,shapes.symbols}

\newtcolorbox{rqsummary}[1]{colback=gray!5,colframe=gray!45,
  title=\textbf{#1},coltitle=black,boxsep=2pt,left=4pt,right=4pt,top=2pt,bottom=2pt}

\usepackage{tabularx}
\usepackage{threeparttable}
\usepackage{multirow}
\usepackage{array}
\usepackage{rotating}

\usepackage{enumitem}

\begin{document}

\title{A Large-Scale Longitudinal Study of Multi-CI Service Adoption}

\author{Taher A. Ghaleb}
\email{taherghaleb@trentu.ca}
\affiliation{
  \institution{Department of Computer Science, Trent University}
  \city{Peterborough}
  \country{Canada}
}

\author{Rasha Bin-Thalab}
\email{r.binthalab@hu.edu.ye}

\affiliation{
  \institution{Computer Engineering Department, Hadhramout University}
  \city{Mukalla}
  \country{Yemen}
}

\affiliation{
  \institution{Information Technology Department, Al-Arab University}
  \city{Mukalla}
  \country{Yemen}
}

\begin{abstract}
Continuous Integration (CI) adoption has evolved from a single practice into a broad landscape of competing services, and many projects no longer rely on a single CI service. However, how this multi-service behavior evolves over a project's lifetime remains unknown. Prior work has examined CI adoption and migration through individual services, specific languages, or developer interviews, leaving the lifecycle of multi-service usage underexplored. In this paper, we conduct a large-scale longitudinal study of multi-CI service adoption covering \num{135227} GitHub repositories from seven programming languages and involving eight CI services. We characterize how CI services are adopted, evolved, and abandoned over time through quantitative and qualitative analyses. About one in five repositories use multiple CI services, often transitional during migration rather than for sustained parallel use. CI service choice is the dominant factor across adoption, complexity, maintenance, and engagement, with language effects limited to specific cases. Our results reveal that the shift to GitHub Actions was a synchronized platform-level transition, crossing 50\% of first adoptions in 2020 across all studied languages, while the decline of Travis CI following its pricing change became the strongest abandonment signal. Our analysis of CI configuration rationales shows that activity is mostly reactive: only about 11\% of commits state a strategic reason, indicating that commit messages capture operational maintenance patterns more reliably than strategic intent. Finally, we find that multi-CI adoption is primarily associated with project maturity, whereas abandonment is more strongly associated with service identity than repository characteristics. These findings provide empirical evidence for CI tooling, migration support, and future research on managing evolving software delivery infrastructure.
\end{abstract}



\maketitle

\section{Introduction}
\label{sec:intro}

Continuous Integration (CI) is a foundational practice in modern software development, automating the building, testing, and validation of code changes so that integration problems surface early rather than at release time~\cite{fowler2006continuous,hilton2016usage}. What started as a single practice has evolved into a crowded landscape of competing services. GitHub Actions, Travis CI, CircleCI, GitLab CI, AppVeyor, Azure Pipelines, Bitbucket Pipelines, and Cirrus CI each offer their own configuration syntax, execution environment, and integration model, and projects must choose among them under real uncertainty about cost, capability, and longevity~\cite{rostami2023usage}.

A project's relationship with these services is rarely a one-time choice. Many repositories run several services at once, migrate from one to another as their needs and the market change, and leave older configurations behind long after they stop using them. Each additional service has its own configuration files, which must be kept current as the platform and the project evolve. Combining or switching services therefore adds maintenance overhead that a single-service setup never incurs~\cite{gallaba2022lessons}. These dynamics are plausibly shaped by the programming language a project is written in, since languages differ in how they build, test, and package code. A systems language producing cross-platform binaries places different demands on CI than an interpreted language whose pipeline centers on dependency validation and packaging~\cite{kiss2022explorative}.

Prior work has studied CI adoption and migration, but largely as static snapshots or through a single lens. Some studies examine one service in isolation, such as the eight-year analysis of CircleCI builds by Gallaba et al.~\cite{gallaba2022lessons}. Others study a single language or package community, such as the co-adoption and migration of services across npm packages~\cite{golzadeh2022rise}, or rely on developer interviews to understand the reasons behind CI decisions~\cite{rostami2023usage}. Our recent study~\cite{chopra_2025} provided an initial analysis of multi-CI adoption and maintenance, but it was limited to Java projects, and the strategic decisions underlying CI service adoption, co-adoption, maintenance, and abandonment remain largely unexplored.

In this paper, we conduct a large-scale longitudinal analysis of multi-CI service adoption, covering seven programming languages, namely C, C++, Go, Java, Python, Ruby, and Rust, and \num{135227} GitHub repositories that adopted at least one of eight CI services. Motivated by prior observations from the earlier study~\cite{chopra_2025}, we investigate the full lifecycle of multi-CI usage through four research questions (RQs), connecting what projects do and when they do it, obtained by mining configuration histories, with why they do it and which projects are involved, as follows:

\begin{itemize}
    \item RQ1 examines how services are adopted and how their configurations evolve in terms of complexity and maintenance effort.
    
    \item RQ2 examines how they are abandoned, switched, and maintained by developers over time.
    
    \item RQ3 investigates why developers adopt, co-adopt, migrate, or abandon them by extracting the natural-language rationale in CI configuration commits.
    
    \item RQ4 examines which repository characteristics predict multi-CI adoption and abandonment.
\end{itemize}

Our results reveal that multi-CI adoption is a common practice, with about one in five repositories running more than one CI service, most often as part of a migration. The composition of this practice is remarkably consistent across languages. Most projects ultimately rely on a single service, and GitHub Actions crossed 50\% of first adoptions in 2020 and reached 96.4\% by 2024. This transition occurred in the same year across all seven languages, indicating a platform-level shift rather than a language-specific one. We also find that CI service choice, rather than programming language, is the dominant factor across adoption, maintenance effort, configuration complexity, and developer engagement. Language effects are limited to specific and interpretable cases such as AppVeyor's concentration in Windows-oriented C, C++, and Rust projects. Our analysis further reveals that the CI service lifecycle was strongly shaped by the decline of Travis CI following its 2020 shift to a paid model, which represents the strongest abandonment signal in the data and appears across all studied languages. Examining the reasons behind these changes, we find that CI configuration activity is mostly reactive: only about 11\% of commits that state a rationale correspond to strategic decisions to adopt, migrate, co-adopt, or abandon a service, while most address deprecations, breakage, and resource constraints. Finally, our models show that multi-CI adoption is primarily associated with project maturity. Repository age and recent activity account for most of the predictive signal. In contrast, abandonment is more strongly associated with service identity than repository characteristics, with Travis CI and CircleCI showing the highest abandonment hazards.

\smallskip\noindent\textbf{Contributions.} This paper makes the following contributions.

\begin{enumerate}
  \item We present a large-scale longitudinal study of multi-CI service adoption spanning seven programming languages, eight CI services, and 135K GitHub repositories, characterizing co-adoption, migration, maintenance, and abandonment over project lifetimes.
  
  \item We establish evidence that CI service choice, rather than programming language, is the dominant factor shaping adoption, configuration complexity, maintenance, and engagement, and identify the transition to GitHub Actions as a synchronized, platform-level shift rather than a language-driven change.
  
  \item We characterize the reasons behind CI decisions from configuration-commit messages using a triangulated design combining manual coding, LLM consensus classification, and topic modeling, revealing that CI activity is predominantly reactive maintenance rather than strategic decision-making.
  
  \item We develop predictive models of multi-CI adoption and abandonment, demonstrating that adoption is primarily associated with project maturity while abandonment is more strongly associated with service identity.
  
  \item We provide a publicly available replication package~\cite{our_replication_package} containing the data, analysis scripts, and detailed results, and qualitative artifacts needed to reproduce and extend our findings.
\end{enumerate}

\smallskip\noindent\textbf{Paper organization.}
The rest of this paper is structured as follows.
Section~\ref{sec:background} provides background on the CI process and the eight services we study.
Section~\ref{sec:related} reviews related work and positions our study.
Section~\ref{sec:design} describes the study design, dataset, and metrics.
Section~\ref{sec:empirical_results} presents our empirical analysis and results.
Section~\ref{sec:discussion} discusses the implications of our findings.
Section~\ref{sec:threats} discusses the threats to validity.
Finally, Section~\ref{sec:conclusion} concludes the paper and outlines future work.

\section{Background}
\label{sec:background}

\subsection{Continuous Integration (CI)}
Continuous Integration (CI) is a software development practice in which code changes are frequently integrated into a shared repository and automatically built and tested to identify integration issues early in the development process~\cite{fowler2006continuous}. In practice, projects enable CI by adding one or more configuration files that a CI service interprets to define and execute a pipeline. When a change is submitted, the CI service typically checks out the code, installs dependencies, builds the project, runs tests, and reports the results to developers. These automated workflows help detect failures early and provide continuous feedback on the correctness of code changes.

\subsection{CI Services}
There are many CI services available for open source projects, each of which has its own configuration format, execution environment, platform integration, and supported operating systems. To capture this diversity, we consider in our study eight commonly used CI services on GitHub:

\begin{itemize}

        \item \textbf{GitHub Actions:} a CI service integrated into GitHub that enables developers to automate workflows directly from their repositories. It supports custom pipelines for building, testing, and deploying code through YAML-based configuration files.

        \item \textbf{Travis CI:} a cloud-based CI service that automates the building and testing of code changes, primarily for open-source projects. It integrates with GitHub and uses a simple \texttt{.travis.yml} file to define build steps.

        \item \textbf{CircleCI:} a CI service that supports fast, scalable pipelines for building, testing, and deploying software. It offers flexible configuration and integrates with various VCS systems, including GitHub and Bitbucket.

        \item \textbf{GitLab CI:} a built-in CI solution within the GitLab ecosystem that automates software development workflows. It uses \texttt{.gitlab-ci.yml} files to define pipelines and supports full integration with GitLab repositories.

        \item \textbf{Bitbucket Pipelines:} a CI service built into Bitbucket Cloud that enables teams to automatically build, test, and deploy code using a configuration file named \texttt{bitbucket-pipelines.yml}. It offers seamless integration with Bitbucket repositories and supports container-based builds using Docker images.
        
        \item \textbf{AppVeyor:} a CI service tailored for Windows-based development, though it also supports Linux and macOS environments. It integrates with popular version control systems such as GitHub, Bitbucket, and GitLab. Configuration is typically done using an \texttt{appveyor.yml} file to specify build and test instructions.

        \item \textbf{Azure Pipelines:} a CI service from Microsoft that supports building, testing, and deploying code across multiple platforms and languages. It integrates with Azure DevOps, GitHub, and other repositories, and allows pipeline definitions through YAML files or a visual editor. It supports both cloud-hosted and self-hosted agents.

        \item \textbf{Cirrus CI:} a flexible CI service designed for modern development workflows. It supports a variety of platforms including Linux, Windows, macOS, and FreeBSD. Cirrus CI integrates with GitHub and uses a \texttt{.cirrus.yml} file to define tasks for building, testing, and deploying applications.
\end{itemize}

A project may relate to these services in ways that go beyond a single static choice, and we use the following terms consistently throughout the paper. A repository exhibits \emph{multi-CI adoption} when it configures two or more distinct CI services at any point in its history. \emph{Co-adoption} refers to running multiple services concurrently, whether to cover complementary platforms or as a hedge, or as a byproduct of an in-progress migration. \emph{Switching}, or sequential adoption, refers to adopting one service after another over time, which we detect from the order of first-configuration commits. A service is \emph{abandoned} in a repository when its configuration file was historically present but no longer exists in the current snapshot, and \emph{obsolete} when its configuration remains present but has not been touched for a long period while the repository otherwise stays active. These lifecycle outcomes, along with the reasons and repository characteristics behind them, drive our four research questions.

\section{Related Work}
\label{sec:related}

We organize prior work along four axes that together frame our study: the adoption, benefits, and challenges of CI; the evolution, migration, and co-usage of CI services; the maintenance and cost of CI configurations; and the influence of programming languages. For each, we state what the closest work established and where our study departs from it.

\subsection{CI Adoption, Benefits, and Challenges}
Early empirical work established that CI is widely used but unevenly so. Hilton et al.~\cite{hilton2016usage} analyzed thousands of GitHub projects and surveyed developers, finding that a large share of projects did not use CI at all, often citing limited experience and setup difficulty, and that CI both raised confidence and slowed some workflows. Elazhary et al.~\cite{elazhary2021uncovering} revisited these trade-offs across teams and reported that the benefits of CI are contingent on how faithfully its practices are followed. The friction of running CI in practice has itself been a recurring subject: Widder et al. documented the pain points that lead projects to leave Travis CI~\cite{widder2018} and later replicated these concerns across a broader Travis population~\cite{widder2019}, identifying configuration difficulty, build duration, and troubleshooting as persistent obstacles. These studies characterize CI as a single practice or a single service at a given point in time. Our study instead follows how the choice among many services plays out across a project's history and across programming languages.

\subsection{CI Service Evolution, Migration, and Co-usage}
The work closest to ours examines how projects move between CI services. Golzadeh et al.~\cite{golzadeh2022rise} studied \num{91810} GitHub repositories of active npm packages and quantitatively traced the co-usage and migration of seven CI services, providing early statistical evidence of the rise of GitHub Actions and the decline of Travis CI within the npm ecosystem. Mazrae et al.~\cite{rostami2023usage} approached the same phenomenon qualitatively, interviewing practitioners to understand the reasons for co-using and migrating between CI services, and reporting the Travis CI to GitHub Actions shift as the dominant migration driven by integration, reliability, and cost. Gallaba et al.~\cite{gallaba2022lessons} analyzed eight years of operational data from a single service, CircleCI, showing how instability and misconfiguration shape build outcomes. A prior study~\cite{chopra_2025} initiated this line of inquiry by studying CI service adoption, complexity, and maintenance in Java projects on GitHub. The present study provides a broader investigation of CI service evolution: it spans seven languages rather than one, quantifies the full lifecycle of CI usage over time rather than focusing on a single-service view or interview-based explanations, and addresses two previously unexplored questions at scale: the stated reasons behind CI decisions extracted from commit messages and the repository characteristics that predict adoption and abandonment.

\subsection{CI Configuration Maintenance, Complexity, and Cost}
A separate body of work treats the CI configuration itself as an evolving, costly artifact. Zampetti et al.~\cite{zampetti2021ci} built a taxonomy of pipeline restructuring activities from commit history and showed that some pipeline components change far more often than others. Valenzuela-Toledo and Bergel~\cite{valenzuela2022} tracked how GitHub Actions workflows evolve after their introduction, and Bouzenia and Pradel~\cite{bouzenia2024resource} quantified the computational resource cost of GitHub Actions workflows and the optimization opportunities left unused. These studies deepen the understanding of one service's configuration, most often GitHub Actions. We complement them by measuring configuration complexity and maintenance effort comparatively across all eight services and seven languages, which enables an understanding of whether the burden they document is a property of the service or of the community in which it is used.

\subsection{CI and Programming Languages}
Programming languages differ in how they build, test, and package code, and a few studies have looked at CI through that lens. Kiss~\cite{kiss2022explorative} examined managed CI usage among open-source C and C++ projects, which are known to have heavier and more platform-dependent build requirements. Prior studies of build duration and build failures~\cite{ghaleb2019empirical,ghaleb2019studying,ghaleb2022interplay} showed that configuration choices have strong associations with CI outcomes, and a recent study of CI configuration in Android apps~\cite{ghaleb2026androidci} found service-specific configuration patterns in a mobile setting. Rather than studying one language in isolation, we study directly, and at scale, whether the programming language or the CI service is the stronger determinant of how CI is adopted, maintained, and abandoned, and we find that the service dominates in nearly every dimension we measure.

\subsection{Positioning}
Table~\ref{tab:related_works_comparison} places our study against the most closely related work along the dimensions that define it. Prior studies each cover part of this space, whether longitudinal evolution~\cite{golzadeh2022rise}, migration reasoning~\cite{rostami2023usage}, or configuration maintenance~\cite{zampetti2021ci,valenzuela2022}, but none combines a cross-language, longitudinal, quantitative account of multi-CI adoption with the stated reasons behind CI decisions and a predictive model of which repositories adopt and abandon services.

\begin{table}[htbp]
\centering
\caption{Positioning of this study against closely related work. \checkmark~= addressed; $\approx$~= partially addressed; blank = not addressed.}
\label{tab:related_works_comparison}
\resizebox{\linewidth}{!}{
\begin{tabular}{llcccccc}
\toprule
\textbf{\shortstack{Study\\~}} & \textbf{\shortstack{Focus\\~}} & \textbf{\shortstack{Longitudinal\\~}} & \textbf{\shortstack{Multi-CI\\~}} & \textbf{\shortstack{Migration\\~}} & \textbf{\shortstack{Cross\\Language}} & \textbf{\shortstack{Reasons\\~}} & \textbf{\shortstack{Predictive\\Model}} \\
\midrule
Hilton et al.~\cite{hilton2016usage}       & CI usage and benefits            & $\approx$ &          &          &          & $\approx$ &          \\
Widder et al.~\cite{widder2018,widder2019} & Travis CI pain points            & $\approx$ &          & $\approx$ &          & \checkmark &          \\
Gallaba et al.~\cite{gallaba2022lessons}   & CircleCI operational data        & \checkmark &          &          &          &          &          \\ Mazrae et al.~\cite{rostami2023usage} & CI/CD co-usage and migration   & $\approx$ & \checkmark & \checkmark &          & \checkmark &          \\
Golzadeh et al.~\cite{golzadeh2022rise}    & CI rise and fall (npm)           & \checkmark & \checkmark & \checkmark &          &          &          \\
Zampetti et al.~\cite{zampetti2021ci}      & Pipeline restructuring           & \checkmark &          &          &          & $\approx$ &          \\
Chopra \& Ghaleb~\cite{chopra_2025}     & Multi-CI in Java                 & \checkmark & \checkmark & \checkmark &          &          &          \\
\midrule
\textbf{This study} & \textbf{Multi-CI across languages} & \checkmark & \checkmark & \checkmark & \checkmark & \checkmark & \checkmark \\
\bottomrule
\end{tabular}
}
\end{table}

\section{Study Design}
\label{sec:design}

\begin{figure*}[t]
\centering
\resizebox{\textwidth}{!}{
\begin{tikzpicture}[
  font=\normalsize,
  db/.style={
    cylinder, shape border rotate=90, draw, thick,
    minimum width=2.0cm, minimum height=1.25cm, aspect=0.32,
    align=center, inner sep=2pt, fill=gray!6
  },
  proc/.style={
    draw,
    thick,
    minimum width=2.9cm,
    minimum height=1.35cm,
    align=center,
    inner sep=3pt,
    fill=white,
    font=\normalsize
  },
  rq/.style={
    draw,
    thick,
    minimum width=3.55cm,
    minimum height=2.1cm,
    align=center,
    inner sep=3pt,
    fill=white,
    font=\normalsize
  },
  band/.style={
    draw, thick, minimum width=16.2cm, minimum height=1.25cm, font=\normalsize,
    align=center, inner sep=4pt, fill=gray!4
  },
  stackback/.style={draw, thin, fill=gray!12},
  ann/.style={font=\normalsize\itshape, align=center},
  arr/.style={-{Stealth[length=5pt]}, thick},
  arrlb/.style={font=\normalsize, align=center}
]

\node[db] (gh) at (0,0) {GitHub\\repositories};
\node[proc, right=0.9cm of gh] (select) {
Repository\\
selection\\[1pt]
active, $\geq 5$ stars,\\
non-fork
};
\node[proc, right=0.7cm of select] (detect) {
CI service\\
detection:\\[1pt]
8 services, from\\
config-file patterns
};
\node[proc, right=0.7cm of detect] (mine) {
History\\
mining\\[1pt]
commits, YAML,\\
contributors, metadata
};

\draw[arr] (gh) -- (select);
\draw[arr] (select) -- (detect);
\draw[arr] (detect) -- (mine);

\node[ann, above=2pt of gh] {\textbf{7 languages:}\\ C, C++, Go, Java,\\ Python, Ruby, Rust};

\node[band] (derived) at ($(gh.south)!0.5!(mine.south)+(0,-1.35)$)
  {\textbf{Study Dataset.}~~\num{135227} repositories across 7 languages and 8 CI services, with full commit and YAML
   configuration histories, contributors, and metadata\\[1pt]
   \num{166184} repository-service pairs (adoption and abandonment lifecycle) ~$\cdot$~ \num{45709} explanatory CI-configuration commits};

\draw[arr] (mine.south) -- (mine.south|-derived.north);

\node[rq, below left=0.8cm and 0.35cm of derived.south] (rq2) {
\textbf{RQ2}\\[1pt]
Abandonment,\\
switching \& survival\\[3pt]
Kaplan--Meier,\\
Cox proportional hazards
};
\node[rq, left=0.7cm of rq2] (rq1) {
\textbf{RQ1}\\[1pt]
Adoption, complexity\\
\& maintenance\\[3pt]
Kruskal--Wallis,\\
effect sizes
};
\node[rq, below right=0.8cm and 0.35cm of derived.south] (rq3) {
\textbf{RQ3}\\[1pt]
Reasons behind\\
CI decisions\\[3pt]
Manual coding, LLM\\
consensus, BERTopic
};
\node[rq, right=0.7cm of rq3] (rq4) {
\textbf{RQ4}\\[1pt]
Predicting adoption\\
\& abandonment\\[3pt]
Logistic regression,\\
SHAP
};

\begin{pgfonlayer}{background}
\foreach \n in {rq1,rq2,rq3,rq4}{
  \draw[stackback] ([xshift=5pt,yshift=5pt]\n.south west) rectangle ([xshift=5pt,yshift=5pt]\n.north east);
  \draw[stackback, fill=gray!5] ([xshift=2.5pt,yshift=2.5pt]\n.south west) rectangle ([xshift=2.5pt,yshift=2.5pt]\n.north east);
}
\end{pgfonlayer}

\draw[arr] (rq1.north|-derived.south) -- (rq1.north);
\draw[arr] (rq2.north|-derived.south) -- (rq2.north);
\draw[arr] (rq3.north|-derived.south) -- (rq3.north);
\draw[arr] (rq4.north|-derived.south) -- (rq4.north);

\end{tikzpicture}
}
\Description{Workflow diagram of the study. GitHub repositories in seven programming languages are filtered to active, non-fork projects. Configuration-file patterns identify eight CI services, and repository metadata, commit histories, configuration histories, and contributors are mined. These data produce three datasets: a repository dataset, repository-service lifecycle pairs, and CI-configuration commits with explanatory messages. The datasets feed four research questions: RQ1 analyzes adoption, configuration complexity, and maintenance; RQ2 analyzes abandonment, migration, and developer engagement; RQ3 analyzes commit-message rationales using manual coding, LLM consensus classification, and topic modeling; and RQ4 models multi-CI adoption and abandonment using logistic regression and SHAP explanations.}
\caption{Overview of the study design, data collection pipeline, and analyses for RQ1--RQ4.}
\label{fig:overview}
\end{figure*}
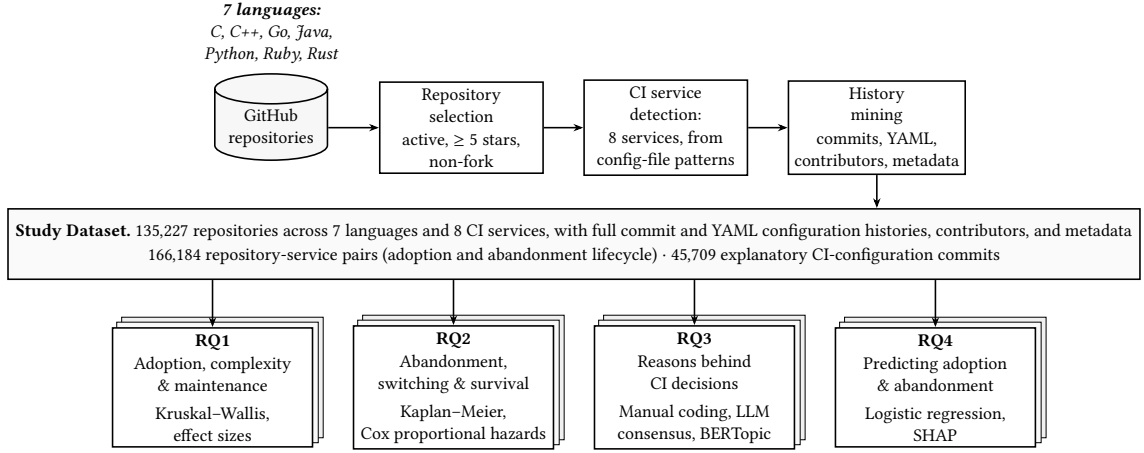

Our study analyzes the CI adoption history of open-source GitHub projects written in seven programming languages, namely C, C++, Go, Java, Python, Ruby, and Rust. We selected these languages to span both compiled systems languages (C, C++, Go, Rust) and interpreted or dynamically typed languages (Python, Ruby), along with Java as the most studied language in CI research, allowing potential language-related differences in CI practices to emerge.
Fig.~\ref{fig:overview} gives an overview of the data collection, processing, and analysis pipeline that we detail in this section.

\subsection{Data Collection}
\label{sec:data}
For each language, we identified active, non-archived, non-fork repositories created from January 2008 to December 2024, with at least five GitHub stars, a filter used to exclude toy and inactive projects while keeping the population broad enough to be representative. We then detected CI service adoption by searching each repository for the configuration files and directories associated with the eight CI services in our study, listed in Table~\ref{tab:ci_yml}. A repository was retained if it adopted at least one of these services at any point in its history, and it was tagged with every service whose configuration we found, which allows a repository to be counted as single-service or multi-service accordingly. This yielded a dataset of \num{135227} repositories that adopted at least one CI service across the seven languages. For the predictive modeling, repositories with missing feature values were excluded, leaving a modeling subset of \num{134264} repositories. For each repository, we collected its metadata, its full commit history, the set of contributing developers, and the CI configuration files together with the commits that touched them.

\begin{table}[htbp]
\centering
\caption{CI services studied and the configuration-file patterns used to detect their adoption.}
\label{tab:ci_yml}
\begin{tabular}{p{4cm}l}
\toprule
\textbf{CI Service} & \textbf{Configuration File Pattern} \\
\midrule
GitHub Actions      & \texttt{.github/workflows/*.yml} \\
Travis CI           & \texttt{.travis.yml} \\
CircleCI            & \texttt{.circleci/config.yml} | \texttt{circle.yml} \\
GitLab CI           & \texttt{.gitlab-ci.yml} \\
AppVeyor            & \texttt{.appveyor.yml} | \texttt{appveyor.yml} \\
Azure Pipelines     & \texttt{azure-pipelines.yml} \\
Bitbucket Pipelines & \texttt{bitbucket-pipelines.yml} \\
Cirrus CI           & \texttt{.cirrus.yml} \\
\bottomrule
\end{tabular}
\end{table}

\subsection{Extracted Characteristics}
\label{sec:characteristics}
From the collected data we derived the following characteristics per repository and per adopted CI service, which form the basis of our analyses.
\begin{itemize}
  \item \textbf{Adoption and evolution:} the set of services a repository adopted and the year each service was first configured, used to trace adoption trends over time.
  \item \textbf{Configuration complexity:} the number of YAML lines in a service's configuration files, aggregated at the repository level.
  \item \textbf{Maintenance activity:} the ratio of commits touching a service's configuration files to the repository's total commit count.
  \item \textbf{Developer engagement:} the ratio of unique developers who authored CI-configuration commits to the repository's total number of contributors.
  \item \textbf{Co-adoption and switching:} which services a repository ran together, and the order and timing of first adoptions across services.
  \item \textbf{Abandonment and obsolescence:} whether a service's configuration was removed from the current snapshot (abandoned) or left untouched for over a year while the repository stayed active (obsolete).
  \item \textbf{Decision rationale:} the natural-language reasons developers state in CI-configuration commit messages (RQ3).
  \item \textbf{Repository features:} activity, popularity, and metadata attributes used to model multi-CI adoption and abandonment (RQ4).
\end{itemize}

\subsection{Data Processing}
We developed scripts to process and analyze our data, normalizing service identifiers and collapsing file-level variants onto a single logical service. This maps the two CircleCI configuration paths to one CircleCI identifier and the two AppVeyor filenames to one AppVeyor identifier. Each research question uses methods suited to its data, which we describe in full within the corresponding section: nonparametric group comparisons and effect sizes for RQ1 and RQ2, survival analysis for time-to-abandonment, a triangulated qualitative and automated analysis of commit messages for RQ3, and logistic regression with SHAP-based explanation for RQ4. Throughout, we pair every significance test with an effect size and interpret practical magnitude rather than $p$-values alone, since the large sample sizes in our dataset render even negligible differences statistically significant. Our replication package provides the data, scripts, and full results~\cite{our_replication_package}.

\section{Empirical Analysis and Results}
\label{sec:empirical_results}

\subsection{RQ1: Adoption, Evolution, Complexity, and Maintenance}
\label{subsec:rq1}

\noindent\textbf{Motivation.}
Developers choosing a CI service face options that overlap in function but differ in how much they cost to set up, how elaborate their configurations become, and how much upkeep they demand. Whether these differences are properties of the service itself or of the programming language that adopts it has direct consequences: if language drives them, advice must be language-specific, whereas if the service drives them, it generalizes. This question examines how the eight services differ in adoption, evolution over time, configuration complexity, and maintenance effort, and whether the programming language changes that picture.

\subsubsection{RQ1.1: Distribution of CI Adoption Rates}~

\smallskip\noindent\textbf{Approach.}
We summarized the number of CI services adopted per repository as a horizontal stacked bar chart, one bar per language plus an overall bar, with each bar normalized to 100\% so that the composition of adoption is comparable across languages of different sizes. To test whether the number of services adopted differs across languages, we applied a Kruskal--Wallis test~\cite{kruskal1952use} and report the effect size $\varepsilon^2$ to convey practical magnitude~\cite{tomczak2014}. We then ran pairwise Mann--Whitney tests~\cite{mann1947} between languages with Holm correction~\cite{holm1979}, reporting Cliff's delta~\cite{cliff1993} for each comparison.

\begin{figure}[htbp]
\centering
\includegraphics[width=.9\linewidth]{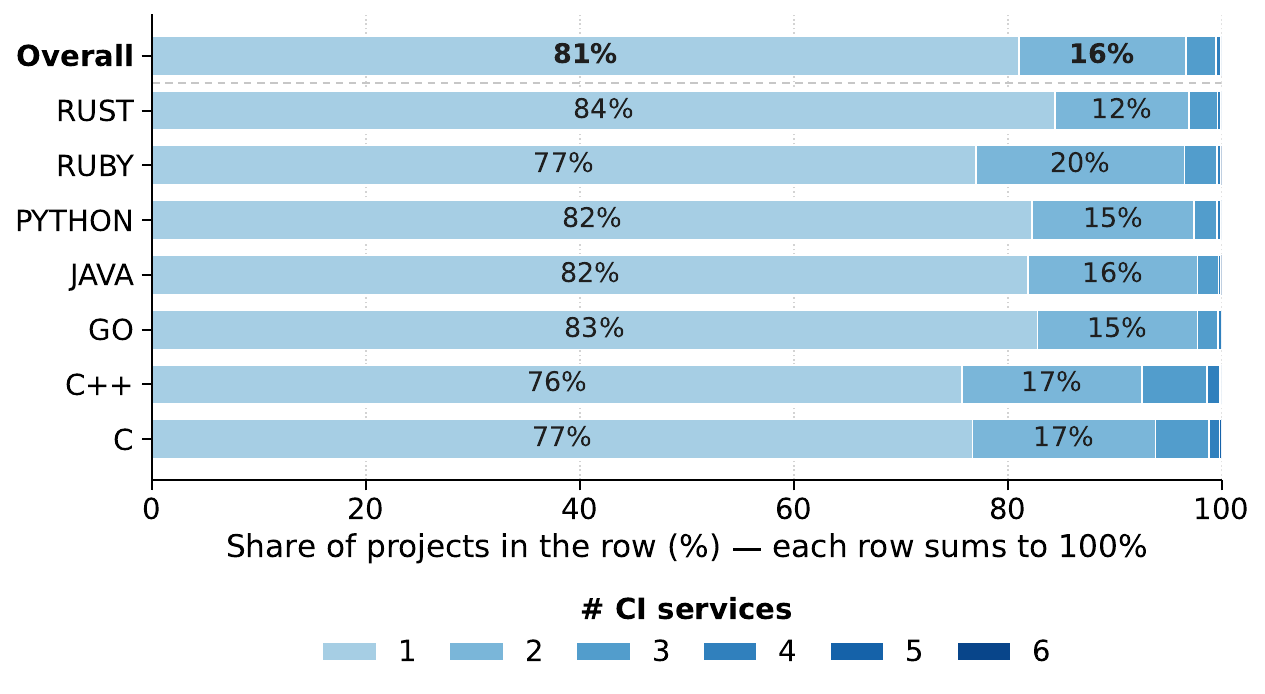}
\Description{Stacked bar chart showing the distribution of repositories by the number of adopted CI services across programming languages. Each bar is normalized to 100\%. Most repositories in every language use a single CI service, a smaller proportion use two services, and only a small fraction use three or more. The overall pattern is similar across languages, though the relative proportions vary slightly.}
\caption{Composition of the number of CI services adopted per repository, by programming language (each bar normalized to 100\%).}
\label{fig:rq1_composition}
\end{figure}

\smallskip\noindent\textbf{Findings.}
The pattern is consistent across all languages: most projects adopt exactly one CI service, ranging from about 77\% for C and Ruby to about 84\% for Rust, with Go, Java, and Python in the low 80s (Fig.~\ref{fig:rq1_composition}). The next most common case is two services, at about 13--20\% of projects, led by Ruby (about 19.5\%) and C (about 17.1\%) and lowest for Rust (about 12.5\%). Three or more services are rare everywhere, tapering to near zero at five or six. Overall, about 81\% of projects adopt a single service and about 16\% adopt two.

The Kruskal--Wallis test is significant ($p \approx 6.3\times10^{-5}$), but the effect size is negligible ($\varepsilon^2 \approx 0.0058$), the expected outcome when a large sample makes a trivial difference detectable. After Holm correction, no pair of languages differs significantly (all adjusted $p = 0.363$), and every Cliff's delta is negligible, the largest in magnitude being about 0.09. The number of CI services a project adopts is therefore, in practical terms, the same across languages.

\smallskip\subsubsection{\textbf{RQ1.2: CI Service Pair Co-occurrence}}~

\smallskip\noindent\textbf{Approach.}
Among repositories running two or more services (\texttt{ci\_services\_count} $\geq 2$), we analyzed which pairs of services co-occur. File-level columns were mapped onto canonical services, collapsing the two CircleCI paths and the two AppVeyor filenames each into one service. For each unordered pair, a repository contributes once if both services are present, and prevalence is the share of a language's multi-CI repositories that exhibit the pair. Every proportion therefore describes composition within the multi-CI population. We tested whether pair prevalence is independent of language with omnibus $\chi^2$ tests on a language by has-pair contingency table. The tests were Holm-adjusted across the ten most frequent pairs, and we report Cram\'{e}r's $V$ as effect size (about 0.10 moderate, at least 0.25 large). To locate the languages driving each association, we computed Haberman-adjusted standardized residuals, treating $|r| > 2$ as a strong contribution. For pairwise language comparisons within each pair, we used two-proportion $z$-tests with Benjamini--Hochberg control~\cite{benjamini1995} across the 21 language pairs. We report the difference in percentage points with 95\% Wald intervals and Cohen's $h$~\cite{cohen1988}. We flag a contrast as actionable only when the false discovery rate is below 0.05 and the difference is at least 3 percentage points. Prevalence alone conflates two effects, namely how often each service is adopted and whether a pair is chosen together more often than independent adoption would predict. We separately computed the co-occurrence ratio of each pair, $P(A \wedge B) / \big(P(A)\,P(B)\big)$, over the full study dataset. Values above one indicate above-expected co-adoption, whereas values below one indicate less frequent co-occurrence than expected under independent adoption.

\begin{figure}[htbp]
\centering
\includegraphics[width=0.9\linewidth]{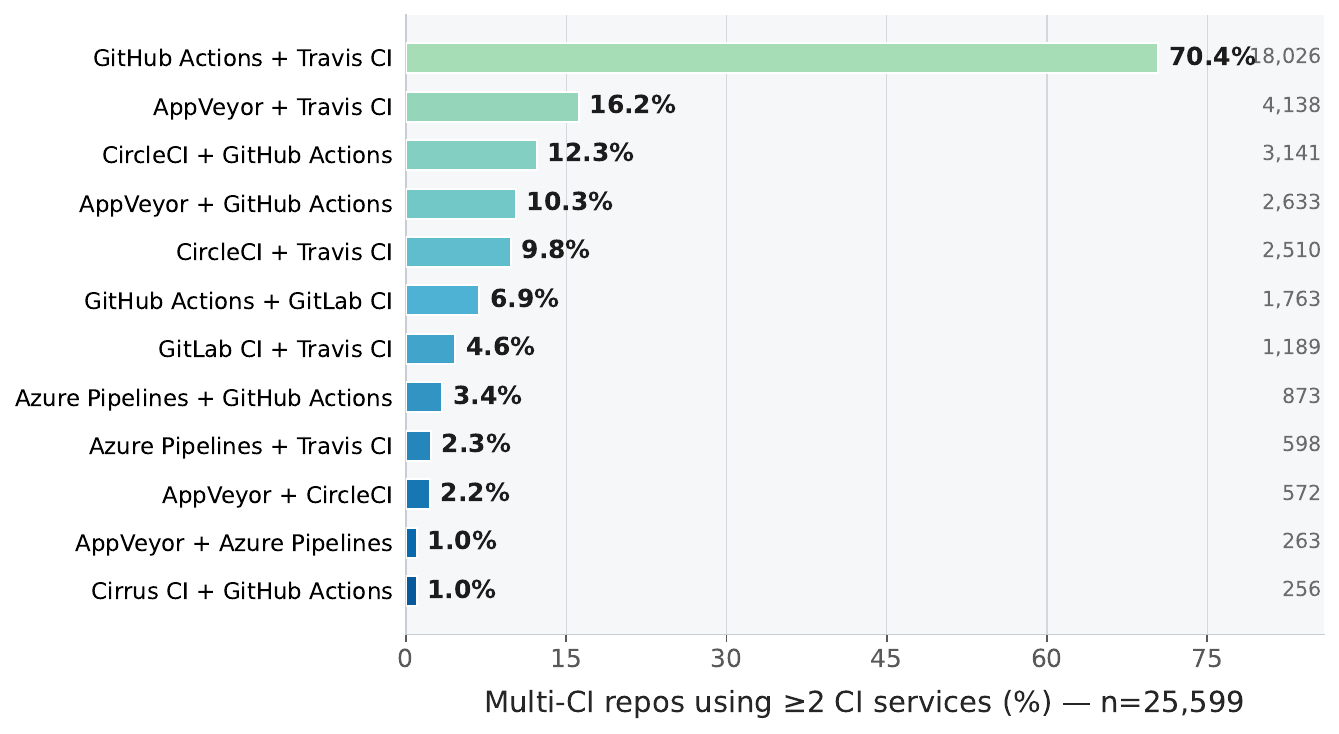}
\Description{Bar chart showing the prevalence of the most common pairs of CI services among repositories that use multiple CI services. The GitHub Actions and Travis CI combination is the most common, followed by GitHub Actions paired with CircleCI and Azure Pipelines. Other service combinations occur less frequently, with a gradual decline in prevalence across the remaining pairs.}
\caption{Prevalence of the most common CI service pairs among multi-CI repositories (n=\num{25599}). Multi-CI repositories have two or more configured services; this snapshot count differs slightly from the \num{25618} repositories identified through first-configuration commit histories in RQ2.2.}
\label{fig:rq1_pairs}
\end{figure}

\begin{figure}[htbp]
\centering
\includegraphics[width=0.9\linewidth]{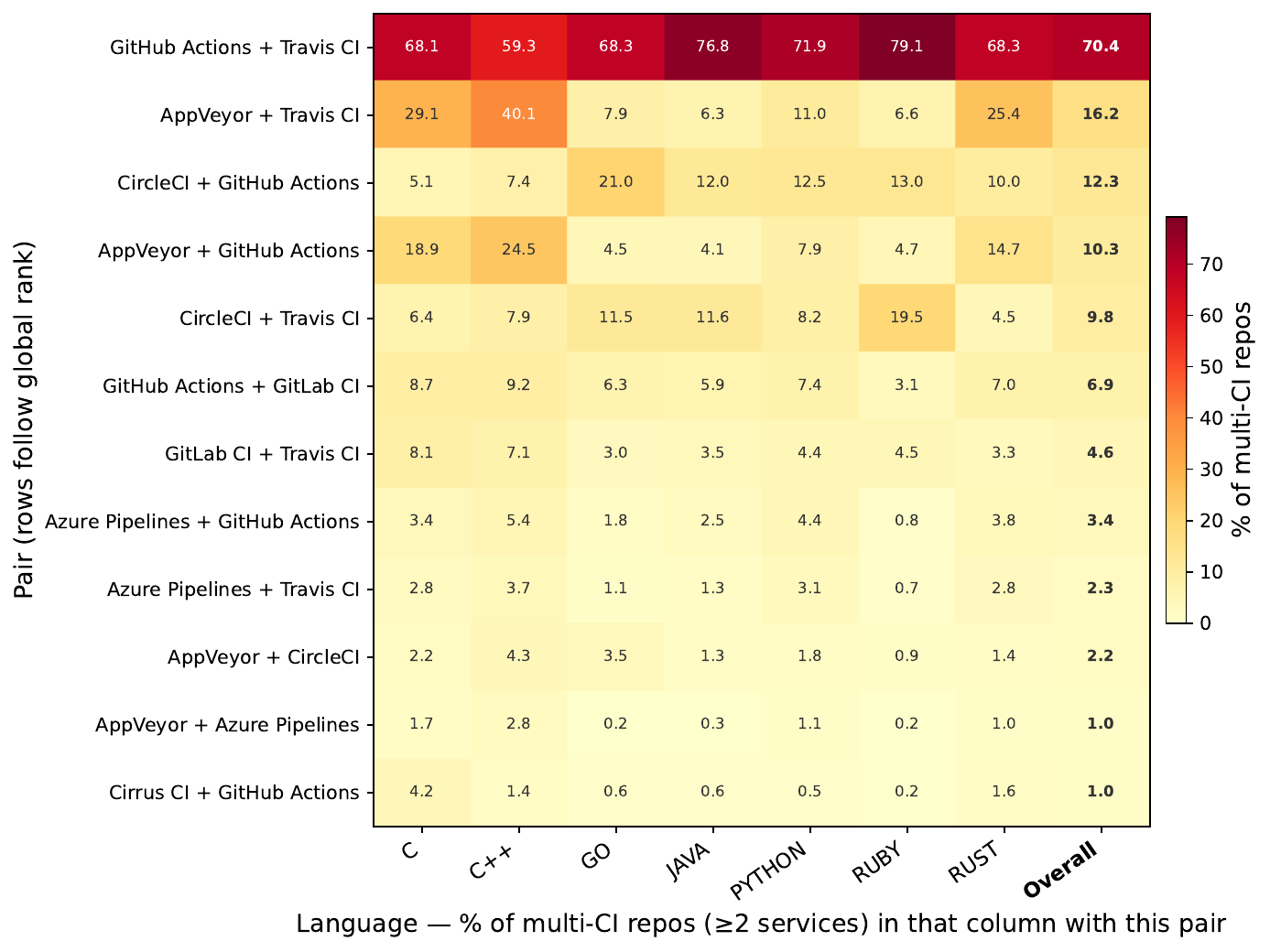}
\Description{Heatmap showing the prevalence of CI service pairs among repositories using at least two CI services, grouped by programming language. Each row represents a CI service pair and each column a programming language. Darker cells indicate a higher percentage of multi-CI repositories in that language using the corresponding pair. The GitHub Actions and Travis CI pair is consistently common across many languages, while other pairs show greater variation in prevalence between languages.}
\caption{Prevalence of CI service pairs among multi-CI repositories ($\geq 2$ services), by language. Each cell is the percentage of that language's multi-CI repositories exhibiting the pair; rows follow global frequency rank.}
\label{fig:rq1_pairs_heatmap}
\end{figure}

\smallskip\noindent\textbf{Findings.}
All ten pairs show significant language dependence after Holm adjustment, but Cram\'{e}r's $V$ spans 0.065--0.321. The strength of the association therefore varies widely (Figs.~\ref{fig:rq1_pairs} and~\ref{fig:rq1_pairs_heatmap}). The clearest and most consistent split separates the native systems languages (C, C++, Rust) from the JVM and scripting languages (Java, Ruby, Python, Go), and it is most apparent in AppVeyor-related pairs. AppVeyor with Travis CI ($V = 0.32$) is the strongest split in the data: C++ leads at 40\%, C at 29\%, and Rust at 25\%, while Java (6.3\%), Ruby (6.6\%), Go (7.9\%), and Python (11\%) cluster at the bottom, with gaps up to 34 percentage points and Cohen's $h \approx 0.86$ for C++ vs. Java or Ruby. AppVeyor with GitHub Actions ($V = 0.23$) reproduces the same structure as projects move off Travis CI, with C++ (24.5\%), C (18.9\%), and Rust (14.7\%) well above Java, Go, and Ruby at 4--5\%. Both patterns reflect the historical reliance of native-language communities on AppVeyor for Windows and multi-platform coverage, a need that rarely arises in JVM or scripting workflows.

Two further service-specific patterns appear in the pairwise results. CircleCI with GitHub Actions ($V = 0.13$) is dominated by Go at 21\%, about double Java, Python, and Ruby, marking it as a distinctly Go combination. CircleCI with Travis CI ($V = 0.13$) instead favors Ruby, which leads at 19.5\% vs. Rust's 4.5\% (15 percentage points, Cohen's $h = 0.49$), consistent with the deep historical roots of Travis and CircleCI in Ruby workflows. GitHub Actions with Travis CI ($V = 0.13$) shows uniformly high co-use (59--79\%), with Ruby and Java still 18--20 points above C++, pointing to projects mid-transition rather than fully moved on. The remaining pairs (GitLab CI and Azure Pipelines combinations, and AppVeyor with CircleCI) are significant but modest ($V = 0.065$ to $0.08$), with directionally consistent but practically small differences.

Accounting for expected co-occurrence reveals a pattern not apparent from raw prevalence alone: most frequent pairs co-occur \emph{less} often than independent adoption would predict. The highest-count pair, GitHub Actions with Travis CI, occurs at only 0.48 times the expected rate (\num{18026} observed vs. \num{37307} expected). CircleCI, GitLab CI, and Azure Pipelines combinations with the two dominant services all sit between 0.58 and 0.85. Services are therefore mostly substitutes, and their co-appearance is the transitional overlap of a migration rather than a deliberate pairing. The clear exceptions all involve AppVeyor: AppVeyor with Azure Pipelines (4.62 times the expected co-occurrence rate), AppVeyor with Travis CI (2.00), and AppVeyor with CircleCI (1.99) are the only strongly complementary combinations, joined by the smaller Cirrus CI with Travis CI (1.38). Concurrent multi-CI use is thus concentrated in the Windows-coverage arrangements that pair AppVeyor with a Linux-oriented service. This is consistent with the systems-language concentration observed in the prevalence split and with the concurrent-combination results in RQ1.6.

\smallskip\subsubsection{\textbf{RQ1.3: Evolution of Adoption Over Time}}~

\smallskip\noindent\textbf{Approach.}
We identified the first year each repository committed to a given service's configuration, yielding one first-adoption event per repository and service, and aggregated these events by year and service. To remove the inflation that dataset growth would otherwise cause in later years, we normalized each year's counts to 100\%, expressing each service's share of that year's first adoptions. We applied this both overall and per language, excluding languages with fewer than 30 repositories in a year to avoid unstable estimates, and used the crossover year, the first year a service exceeds 50\% share, as a simple marker of dominance shifts.

\begin{figure}[htbp]
\centering
\includegraphics[width=0.9\linewidth]{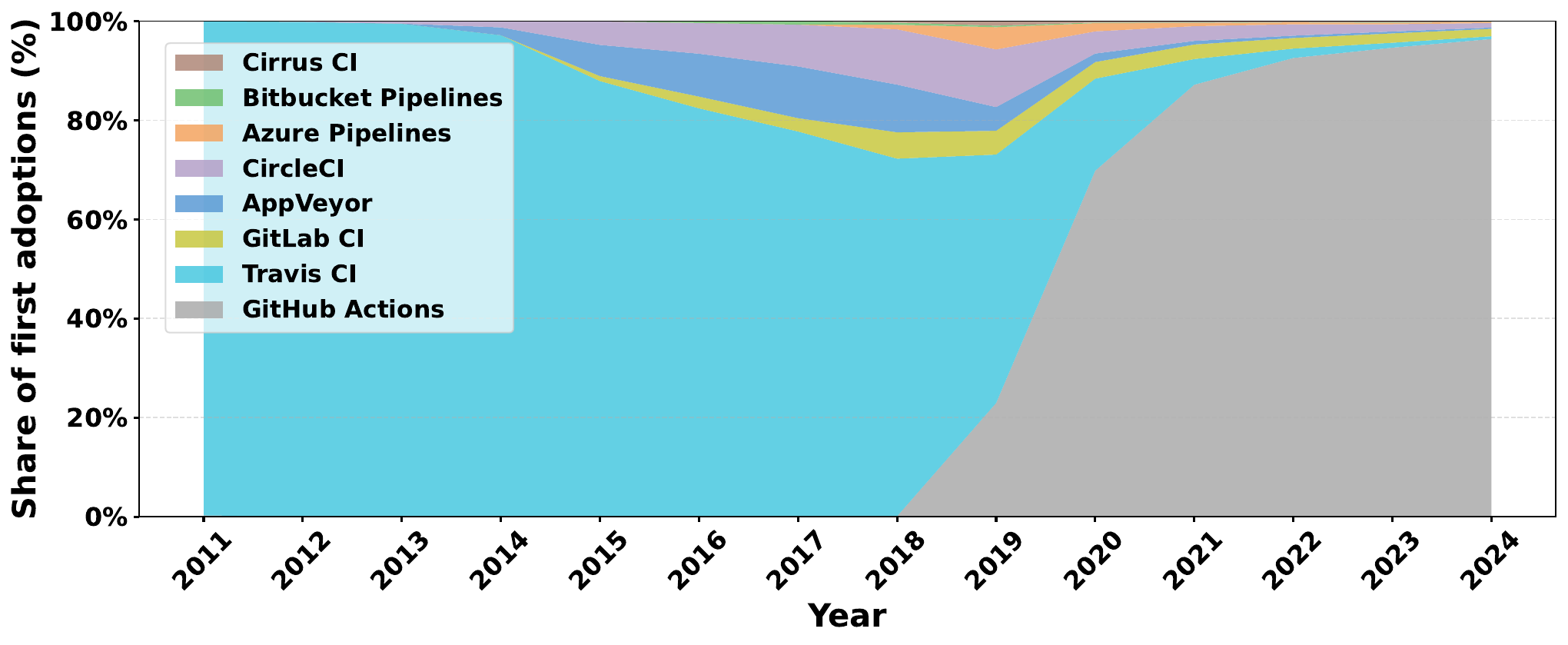}
\Description{Stacked area chart showing the annual share of first-time CI service adoptions from 2011 through 2024 across all programming languages. The composition of adopted services changes over time, with early years dominated by Travis CI and later years increasingly dominated by GitHub Actions. Other CI services account for smaller shares and fluctuate over the study period.}
\caption{Share of first-time CI adoptions per year, by service, across all languages (2011--2024).}
\label{fig:rq1_evolution}
\end{figure}

\smallskip\noindent\textbf{Findings.}
Travis CI held near-total share of first adoptions in the early years, consistent with its status as the first widely adopted service on GitHub, though shares before 2014 rest on small absolute counts and should be read cautiously (Fig.~\ref{fig:rq1_evolution}). GitHub Actions crossed the 50\% first-adoption threshold in 2020, one year after its launch, and reached 96.4\% by 2024, a rapid and near-complete consolidation around one platform. This crossover occurred in 2020 in all seven languages, which points to a platform-level cause, most plausibly the combination of native GitHub integration and the concurrent shift of Travis CI to a paid model, rather than anything language-specific. The main language-specific deviation is AppVeyor, whose first-adoption share peaks at 25.1\% in C++ and 17.5\% in C, vs. 10.4\% overall, matching its Windows-native positioning and the stronger Windows build requirements of systems languages.

\smallskip\subsubsection{\textbf{RQ1.4: Maintenance Activity}}~

\smallskip\noindent\textbf{Approach.}
For each repository, we computed the ratio of commits touching a service's configuration files to the repository's total commit count, yielding one maintenance ratio per repository and service. We aggregated these by service and reported the median, mean, and the 75th and 90th percentiles, treating the median as the primary statistic given the right skew. To test whether maintenance effort differs across languages, we applied a Kruskal--Wallis test per service across the seven languages with Bonferroni correction (adjusted $\alpha = 0.006$), reporting $\eta^2$ as effect size. We excluded 83 repositories for which CI-related commits exceeded the total commit count for at least one service, which indicates noise in the repository-level commit counts.
Considering that automated bots may author a share of CI-configuration commits and could inflate the maintenance ratio, we identified GitHub bot authors and recomputed each service's median ratio with bot commits removed from the numerator. A GitHub author counts as a bot when its login contains GitHub's reserved \texttt{[bot]} suffix, is typed \texttt{Bot} in the contributor table, or matches a curated list of known automation accounts. Given that any one of these three criteria is sufficient, the curated list is not a hard dependency. An automation account missing from it is still flagged by the \texttt{[bot]} suffix or the typed \texttt{Bot} field, and the rule therefore does not understate the bot share. We further analyzed whether maintenance ranking is influenced by two potential confounding factors. To separate maintenance from configuration size, we correlated a repository's YAML line count with its number of configuration-touching commits using Spearman's $\rho$, overall and per service. To separate maintenance from how long a service has been in use, we recomputed activity as commits per active year and commits per 100 configuration lines, and re-ran the cross-service Kruskal--Wallis test on each normalized measure.
\begin{figure}[!t]
\centering
\includegraphics[width=\linewidth]{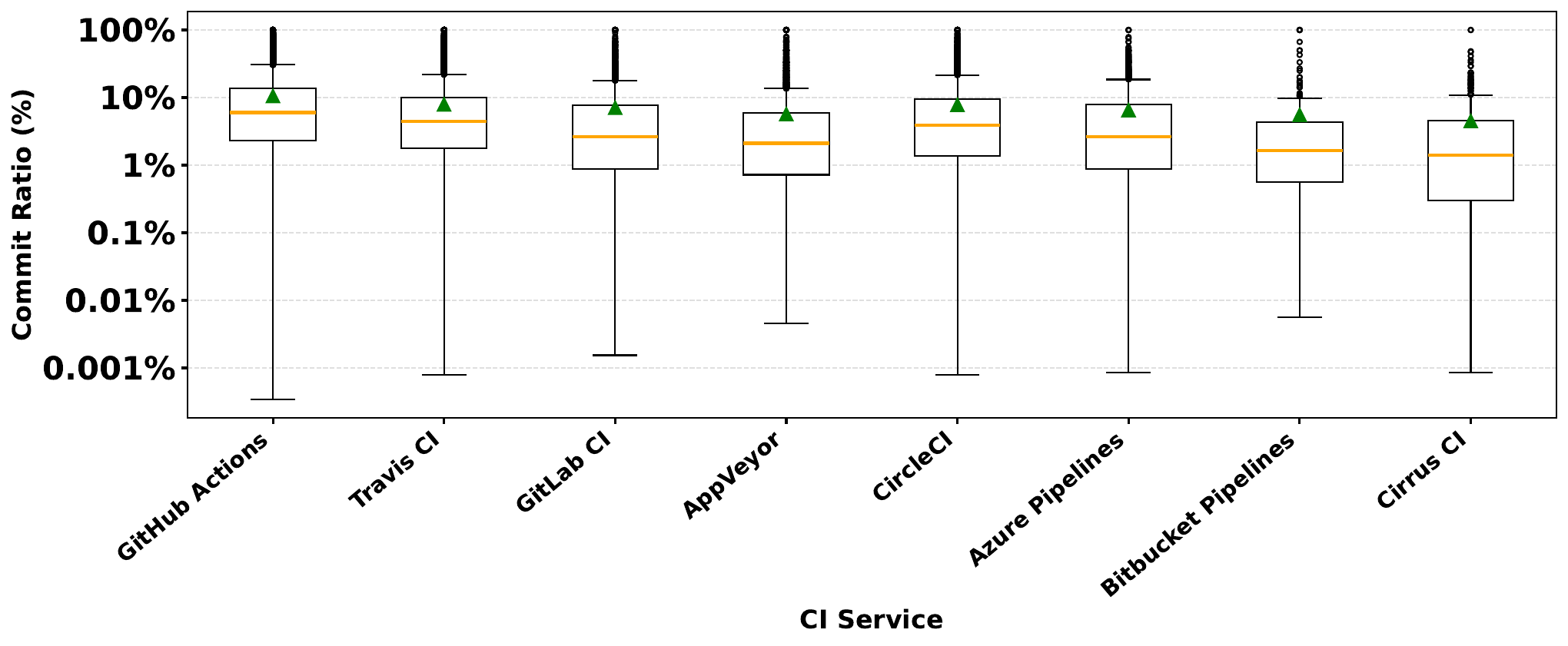}
\Description{Boxplots showing the distribution of CI maintenance ratios, measured as the percentage of commits that modify CI configuration files, for each CI service across all programming languages on a logarithmic scale. Each box shows the interquartile range, the orange horizontal line marks the median, and the green triangle marks the mean. Maintenance ratios vary substantially across services, with long-tailed distributions indicating that while most repositories make relatively few CI-related commits, a smaller number perform much more frequent CI maintenance.}
\caption{Distribution of CI maintenance ratios (commit ratio \%) per service, across all languages (log scale). Orange lines denote medians; green triangles denote means.}
\label{fig:rq1_maintenance}
\end{figure}

\smallskip\noindent\textbf{Findings.}
GitHub Actions has the highest median maintenance ratio at 6.01\%, followed by Travis CI (4.44\%) and CircleCI (3.92\%), while Bitbucket Pipelines (1.66\%) and Cirrus CI (1.42\%) are lowest (Fig.~\ref{fig:rq1_maintenance}). Every service's distribution is heavily right-skewed, with means above medians. A small subset of repositories therefore accounts for a disproportionate share of maintenance activity. Kruskal--Wallis tests are significant for six of the eight services after Bonferroni correction, but the effect sizes are uniformly small or negligible ($\eta^2$ 0.009--0.044), and the programming language has little practical bearing on maintenance effort. The one large value, for Cirrus CI ($\eta^2 = 0.190$), rests on insufficient per-language samples and should not be over-read. Maintenance overhead is thus a service-level property, with GitHub Actions consistently the most maintenance-intensive across all languages.

Three robustness checks confirm that this ranking is not an artifact. Bots author only 6.5\% of the \num{2423030} CI-configuration commits in this maintenance sample (\num{156780}, computed on the repositories retained above). They are concentrated in GitHub Actions (10.1\%) and near zero elsewhere, thus removing them barely moves the medians: GitHub Actions falls from 6.01\% to 5.68\% and remains the clear top of the ranking, while every other service shifts by at most 0.03 percentage points. Configuration size and maintenance churn are moderately correlated (Spearman $\rho = 0.54$ overall, 0.34--0.62 per service, all $p < 0.001$), meaning more elaborate configurations do attract more upkeep. However, the correlation with the size-normalized \emph{ratio} is weak ($\rho = 0.16$ overall), hence service ranking is not merely a restatement of the complexity ranking from RQ1.5. The ranking also survives removing the service-age confounding factors. Normalized to commits per 100 configuration lines, GitHub Actions is in fact the \emph{least} churny per line (8.89), while Travis CI is the most (25.0). The cross-service difference remains large ($\eta^2 = 0.15$). GitHub Actions is thus maintenance-intensive because its configurations are larger and more actively developed, not because each line is edited more often.

\smallskip\subsubsection{\textbf{RQ1.5: Configuration Complexity}}~

\smallskip\noindent\textbf{Approach.}
We measured the number of YAML lines in a service's configuration files, aggregated at the repository level, yielding one complexity value per repository and service, and report the median, mean, and 75th and 90th percentiles with the median as the primary statistic. To test whether complexity differs across services, we applied one Kruskal--Wallis test across all services, followed by pairwise Dunn tests~\cite{dunn1964} with Bonferroni correction to locate the differing pairs. We separately tested whether complexity differs across languages per service with a Bonferroni-corrected Kruskal--Wallis test (adjusted $\alpha = 0.006$), reporting $\eta^2$ throughout.

\begin{figure}[htbp]
\centering
\includegraphics[width=0.9\linewidth]{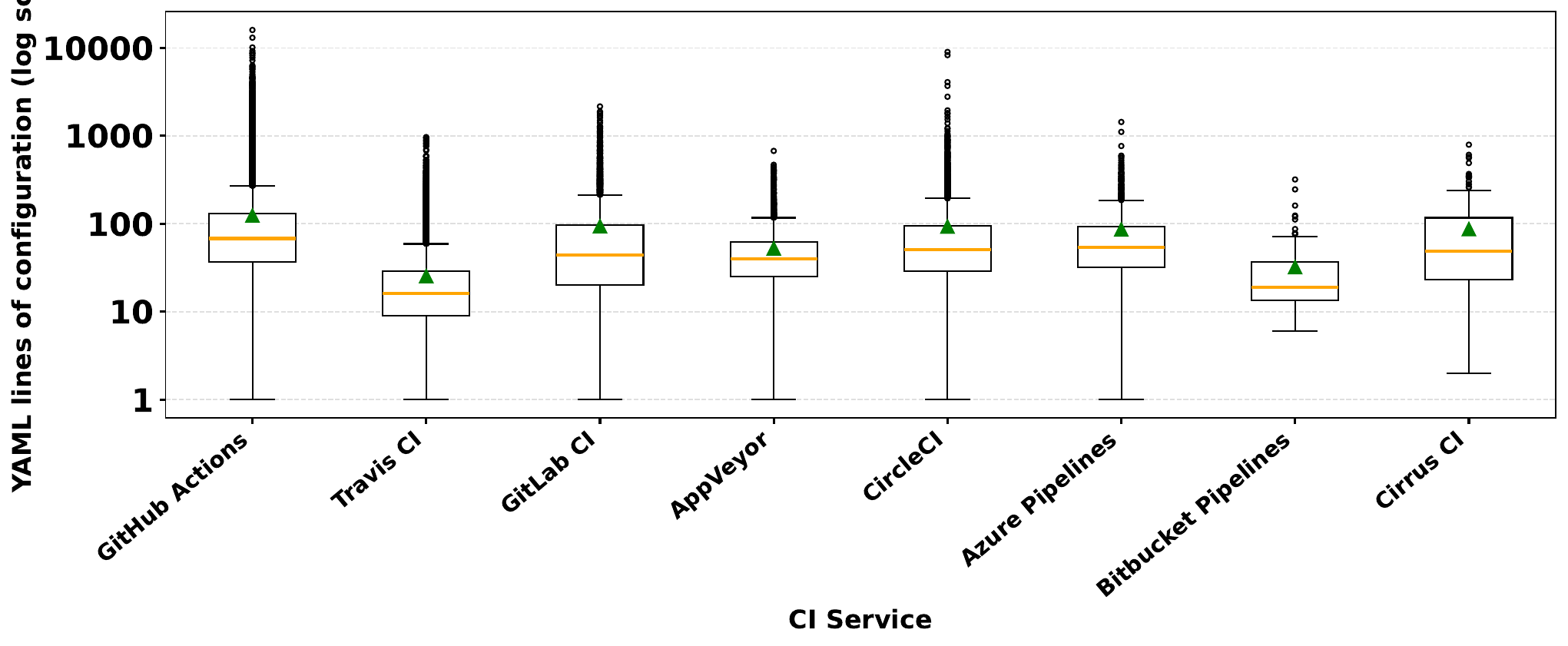}
\Description{Boxplots showing the distribution of CI configuration complexity, measured as the number of lines in YAML configuration files, for each CI service across all programming languages on a logarithmic scale. Each box shows the interquartile range, the orange horizontal line marks the median, and the green triangle marks the mean. Configuration complexity varies across services, with generally right-skewed distributions indicating that most repositories have relatively compact CI configurations while a smaller number maintain substantially larger YAML files.}
\caption{Distribution of YAML configuration complexity (lines) per CI service, across all languages (log scale). Orange lines denote medians; green triangles denote means.}
\label{fig:rq1_complexity}
\end{figure}

\smallskip\noindent\textbf{Findings.}
Services differ substantially in configuration complexity (Kruskal--Wallis $H = 47{,}604.90$, $p < 0.001$, $\eta^2 = 0.35$, large effect; Fig.~\ref{fig:rq1_complexity}). GitHub Actions has by far the highest median at 68 lines, about four times Travis CI (16 lines) and more than three times AppVeyor (40 lines) and Bitbucket Pipelines (19 lines). GitLab CI (44), CircleCI (51), Azure Pipelines (54), and Cirrus CI (49) form a middle band. Pairwise Dunn tests confirm three distinct tiers: GitHub Actions alone at the top; a middle tier of CircleCI, Azure Pipelines, GitLab CI, and Cirrus CI with no significant differences among them; and a low tier of Travis CI and Bitbucket Pipelines, which do not differ from each other ($p = 0.87$).

Service choice remains the dominant driver of complexity within every language (per-language $\eta^2$ 0.16--0.60, all large). Language explains little additional variance for most services, and GitHub Actions in particular shows only a small language effect ($\eta^2 = 0.015$), confirming that its verbosity is intrinsic to the service. The exception is Travis CI, which shows a large language effect ($\eta^2 = 0.186$), driven by minimal Ruby configurations (median 8 lines) vs. C and C++ (medians 24 and 36 lines), where matrix build configurations add length. Azure Pipelines shows a medium language effect ($\eta^2 = 0.074$) that rests on small per-language samples, and Bitbucket Pipelines and Cirrus CI show no significant language differences after correction. Configuration complexity is therefore primarily a matter of service choice, with language playing a secondary role only for Travis CI.

\smallskip\subsubsection{\textbf{RQ1.6: Concurrent Multi-CI Combinations}}~

\smallskip\noindent\textbf{Approach.}
The preceding analyses count any repository that ever adopted two or more services. Here, we restricted attention to services that are simultaneously active in the current snapshot, that is, whose configuration files still exist, to characterize concurrent service use rather than sequential adoption where an earlier service has since been removed. We counted repositories by the exact set of currently active services and report the most common combinations overall and per language.

\smallskip\noindent\textbf{Findings.}
Of the repositories in our dataset, \num{7967} keep two or more CI services simultaneously active, well below the \num{25618} that ever adopted multiple services (RQ2.2), which confirms that much multi-CI adoption is transitional, with the older service eventually removed. Among concurrent combinations, GitHub Actions with Travis CI is by far the most common (\num{2827} repositories, 35.5\% of concurrent multi-CI repositories), followed by AppVeyor with Travis CI (17.8\%), CircleCI with GitHub Actions (9.5\%), and GitHub Actions with GitLab CI (9.3\%). Three-service combinations are uncommon but present, led by AppVeyor with GitHub Actions and Travis CI (3.1\%). The per-language picture mirrors RQ1.2, with AppVeyor combinations far more prevalent in C, C++, and Rust, reinforcing that concurrent co-existence, where it persists, is most often the residue of an in-progress migration onto GitHub Actions or a deliberate Windows-coverage arrangement in systems languages.

\begin{rqsummary}{RQ1 Summary}
CI service choice, not programming language, is the dominant factor across adoption, evolution, complexity, and maintenance. Most projects use a single service, GitHub Actions consolidated first adoptions across all seven languages by 2020 (96.4\% by 2024), and it is both the most complex to configure and the most maintenance-intensive. Language matters only in specific cases, most visibly AppVeyor's concentration in the Windows-oriented C, C++, and Rust projects.
\end{rqsummary}

\subsection{RQ2: Abandonment, Switching, and Developer Engagement}
\label{subsec:rq2}

\noindent\textbf{Motivation.}
Adopting a CI service is only the start of a relationship that may end in the service being replaced or quietly left behind. Understanding when and how services are abandoned, in what order projects move between them, how long they are retained, and how widely their upkeep is shared among contributors reveals the maintenance debt and coordination cost that multi-CI use accumulates over time. This question follows the CI service lifecycle from abandonment and obsolescence through sequential switching, developer engagement, and the timing of abandonment.

\smallskip\subsubsection{\textbf{RQ2.1: Abandonment and Obsolescence}}~

\smallskip\noindent\textbf{Approach.}
We classified a service as \emph{abandoned} in a repository when its configuration file was historically present but no longer exists in the current snapshot. We classified it as \emph{obsolete} when the configuration remains present but has not been touched for more than a year while the repository otherwise has later commits. This marks the service as stale despite ongoing activity. When a repository-service pair qualifies as both, we assigned it to abandoned. We computed abandonment and obsolescence rates per service as the proportion of that service's repository-service pairs in each state, overall and per language, excluding languages with fewer than 30 repositories. We tested whether rates differ across services and languages with chi-squared tests of independence, reporting Cram\'{e}r's $V$, and applied Bonferroni correction (adjusted $\alpha = 0.006$) for the cross-language comparisons.

\smallskip\noindent\textbf{Findings.}
Abandonment and obsolescence differ significantly across services ($\chi^2 = 33{,}337.23$, $p < 0.001$, Cram\'{e}r's $V = 0.32$, medium effect), and the association holds for abandonment alone ($V = 0.40$) and obsolescence alone ($V = 0.32$). Lifecycle outcome is therefore meaningfully tied to which service a project chose (Table~\ref{tab:abandon}). Abandonment rates vary widely: Bitbucket Pipelines (51.7\%), Azure Pipelines (47.0\%), AppVeyor (43.3\%), and CircleCI (42.4\%) sit at the top, with more than four in ten adopting repositories having removed the configuration. Travis CI accounts for the largest absolute number of abandoned repositories (\num{17893}), reflecting both its historically high adoption and its decline after the 2020 shift to a paid model, though its abandonment \emph{rate} (32.6\%) is moderate. GitHub Actions has the lowest abandonment rate (6.6\%), consistent with its recent launch and continued dominance. For obsolescence, AppVeyor stands out with both high abandonment (43.3\%) and high obsolescence (24.6\%), a combined lifecycle loss near 68\%, whereas Travis CI's obsolescence (7.3\%) is low relative to its abandonment, indicating that projects tended to remove Travis CI outright rather than leave it stale.

Across languages, abandonment differs significantly for most services after Bonferroni correction, but the effect sizes are negligible to small (Cram\'{e}r's $V$ 0.064--0.185). GitLab CI, AppVeyor, and Azure Pipelines show the largest language effects ($V = 0.175$, $0.167$, and $0.185$), while GitHub Actions and CircleCI show negligible variation ($V = 0.064$ and $0.087$). Abandonment and obsolescence are thus primarily service-level phenomena, with language a secondary and practically small factor.

\begin{table}[htbp]
\centering
\caption{Abandonment and obsolescence of CI services across all languages. Rates are shares of each service's repository-service pairs.}
\label{tab:abandon}
\begin{tabular}{p{4cm}rrrrr}
\toprule
\textbf{CI Service} & \textbf{Total} & \textbf{Aband. \#} & \textbf{Aband. \%} & \textbf{Obs. \#} & \textbf{Obs. \%} \\
\midrule
GitHub Actions      & \num{91870} & \num{6049}  & 6.6\%  & \num{1231} & 1.3\%  \\
Travis CI           & \num{54947} & \num{17893} & 32.6\% & \num{4020} & 7.3\%  \\
CircleCI            & \num{7652}  & \num{3242}  & 42.4\% & 595        & 7.8\%  \\
AppVeyor            & \num{5087}  & \num{2204}  & 43.3\% & \num{1253} & 24.6\% \\
GitLab CI           & \num{4452}  & \num{1144}  & 25.7\% & 664        & 14.9\% \\
Azure Pipelines     & \num{1511}  & 710         & 47.0\% & 133        & 8.8\%  \\
Cirrus CI           & 396         & 104         & 26.3\% & 65         & 16.4\% \\
Bitbucket Pipelines & 269         & 139         & 51.7\% & 45         & 16.7\% \\
\midrule
\textbf{Overall}    & \num{166184} & \num{31485} & 18.9\% & \num{8006} & 4.8\% \\
\bottomrule
\end{tabular}
\end{table}

\smallskip\subsubsection{\textbf{RQ2.2: Sequential Adoption (Switching)}}~

\smallskip\noindent\textbf{Approach.}
We identified repositories that adopted more than one service over time by detecting the first commit touching each service's configuration files, ordered the services by first-commit date, and recorded every consecutive (earlier, later) pair together with the number of days between the two first-adoption dates. We report the overall multi-service adoption rate, the most common adoption pairs, and the elapsed time between adoptions. This analysis captures temporal ordering of adoption and does not require the earlier service to have been removed, which is assessed separately in RQ2.1. To test whether adoption timing differs across destination services and across languages, we applied Kruskal--Wallis tests with $\eta^2$, using Bonferroni correction ($\alpha = 0.007$) for the cross-language comparison.

\smallskip\noindent\textbf{Findings.}
Of the \num{135227} repositories in our dataset, \num{25618} (18.9\%) adopted more than one service over time, yielding \num{30957} sequential adoption pairs. The overall median time between adoptions is 651 days. Multi-service adoption is generally not an immediate decision: only 23.8\% of adoption pairs occur within three months and 37.5\% within one year, and no repository re-adopted a previously used service. The dominant pair is Travis CI followed by GitHub Actions (\num{14731} pairs, median \num{1141} days), consistent with the mass migration after the 2020 Travis CI pricing change (Table~\ref{tab:top_seq_pairs}). The second most common is Travis CI followed by AppVeyor (\num{2557} pairs, median 170 days), reflecting earlier multi-service experimentation. Two pairs, AppVeyor followed by Travis CI and CircleCI followed by Travis CI, show a median of zero days, indicating services configured together at project inception rather than true sequential adoption.

Adoption timing differs significantly across destination services (Kruskal--Wallis $H = 8{,}791.49$, $p < 0.001$, $\eta^2 = 0.28$, large effect). Which service a project adds is therefore strongly tied to how long it waits. Per-language tests are significant for all seven languages after correction with large effects ($\eta^2$ 0.23--0.39; Table~\ref{tab:kw_languages}). Ruby repositories wait longest before adding a service (median \num{1266} days), while C++ and Rust move quickest (medians 412 and 410 days), suggesting that languages differ in the pace, though not the eventual destination, of CI diversification (Table~\ref{tab:lang_adoption}).

\begin{table}[htbp]
\centering
\caption{Top 10 sequential CI service adoption pairs.}
\label{tab:top_seq_pairs}
\begin{tabular}{p{3cm}p{3cm}rr}
\toprule
\textbf{From} & \textbf{To} & \textbf{Count} & \textbf{Median Days} \\
\midrule
Travis CI       & GitHub Actions & \num{14731} & \num{1141} \\
Travis CI       & AppVeyor       & \num{2557}  & 170        \\
CircleCI        & GitHub Actions & \num{2390}  & 606        \\
Travis CI       & CircleCI       & \num{1569}  & 438        \\
AppVeyor        & GitHub Actions & \num{1563}  & \num{1160} \\
AppVeyor        & Travis CI      & \num{1302}  & 0          \\
GitLab CI       & GitHub Actions & \num{1125}  & 466        \\
Azure Pipelines & GitHub Actions & 719         & 435        \\
Travis CI       & GitLab CI      & 606         & 541        \\
CircleCI        & Travis CI      & 531         & 0          \\
\bottomrule
\end{tabular}
\end{table}

\begin{table}[htbp]
\centering
\caption{Sequential adoption intervals across programming languages.}
\label{tab:lang_adoption}
\begin{tabular}{lrrrr}
\toprule
\textbf{Language} & \textbf{Multi-CI Repos} & \textbf{Median Days} & \textbf{$\leq$90d (\%)} & \textbf{$\leq$365d (\%)} \\
\midrule
C      & \num{1946} & 754        & 23.4 & 35.0 \\
C++    & \num{3522} & 412        & 33.2 & 47.5 \\
Go     & \num{3849} & 651        & 20.7 & 36.2 \\
Java   & \num{3442} & 884        & 18.1 & 30.2 \\
Python & \num{7827} & 608        & 22.7 & 37.7 \\
Ruby   & \num{2651} & \num{1266} & 16.9 & 24.6 \\
Rust   & \num{2381} & 410        & 31.9 & 48.2 \\
\bottomrule
\end{tabular}
\end{table}

\begin{table}[htbp]
\centering
\caption{Kruskal--Wallis tests of adoption intervals across destination services, within each language.}
\label{tab:kw_languages}
\begin{tabular}{p{2cm}rrrl}
\toprule
\textbf{Language} & \textbf{H} & \textbf{p-value} & \textbf{$\eta^2$} & \textbf{Effect} \\
\midrule
C      & 891.41  & $<0.001$ & 0.345 & Large \\
C++    & 1810.16 & $<0.001$ & 0.374 & Large \\
Go     & 1066.42 & $<0.001$ & 0.240 & Large \\
Java   & 920.64  & $<0.001$ & 0.233 & Large \\
Python & 2093.32 & $<0.001$ & 0.228 & Large \\
Ruby   & 776.68  & $<0.001$ & 0.249 & Large \\
Rust   & 1149.78 & $<0.001$ & 0.394 & Large \\
\bottomrule
\end{tabular}
\end{table}

\smallskip\subsubsection{\textbf{RQ2.3: Developer Engagement in CI Configuration}}~

\smallskip\noindent\textbf{Approach.}
For each repository, we computed the ratio of unique authors of CI-configuration commits to the repository's total number of contributors, yielding one author-to-contributor ratio per repository and service. A ratio of 1.0 means every contributor has touched CI configuration, while lower values mean CI upkeep is concentrated in a subset. We excluded \num{108} repository-service pairs where the ratio exceeded 1.0, which indicates a mismatch between data sources, and report the median, mean, and 75th and 90th percentiles, with the median primary. We tested differences across services and languages with Kruskal--Wallis tests and $\eta^2$, applying Bonferroni correction ($\alpha = 0.006$) for the cross-language comparisons. Given that a bot that commits CI configuration would count as an author and could depress the ratio, we recomputed the median with bot authors removed, using the same bot-identification rule as in RQ1.4.

\begin{figure}[t]
\centering
\includegraphics[width=0.9\linewidth]{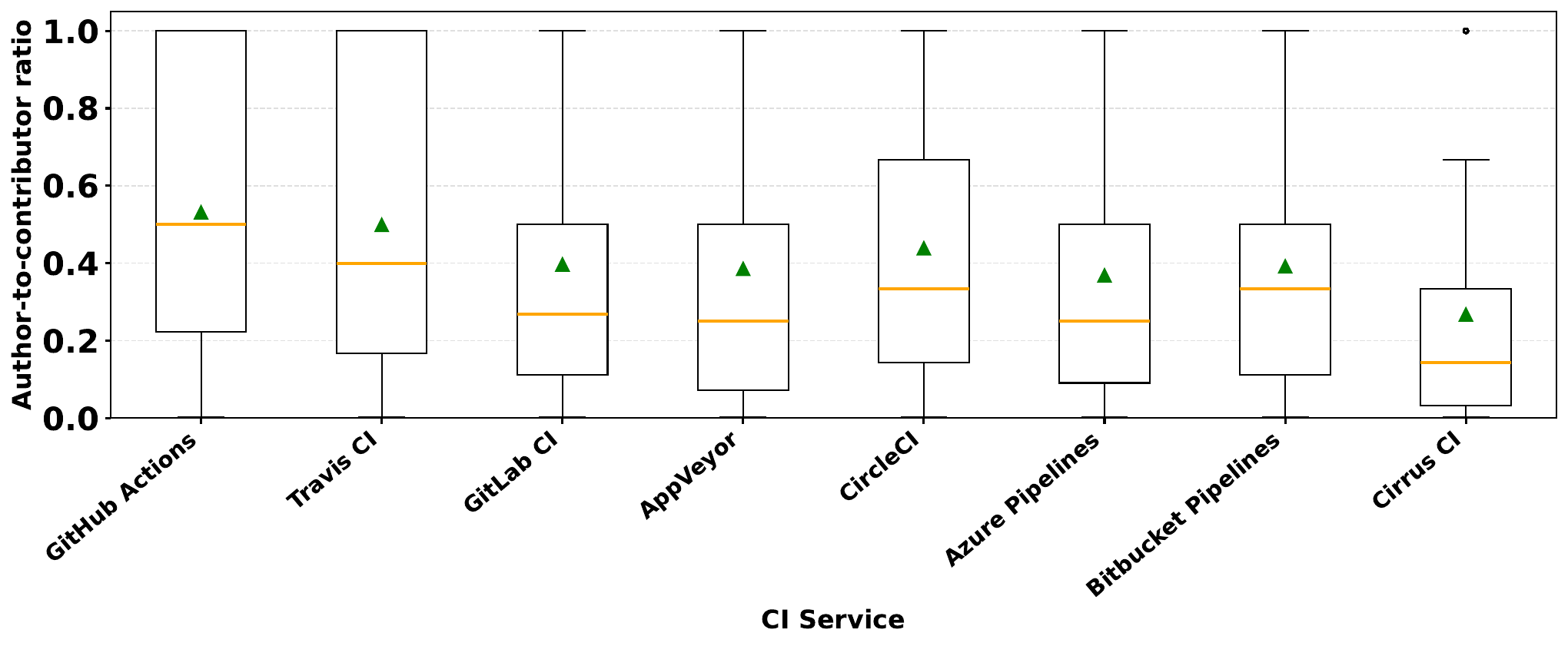}
\Description{Boxplots showing the distribution of author-to-contributor ratios for repositories using each CI service across all programming languages. Each box shows the interquartile range, the orange horizontal line marks the median, and the green triangle marks the mean. The distributions differ across CI services and are generally right-skewed, indicating that while many repositories have relatively low author-to-contributor ratios, some repositories exhibit substantially higher values.}
\caption{Distribution of author-to-contributor ratios per CI service, across all languages. Orange lines denote medians; green triangles denote means.}
\label{fig:rq2_engagement}
\end{figure}

\smallskip\noindent\textbf{Findings.}
GitHub Actions has the highest median engagement ratio (0.50), meaning that in a typical repository half of the contributors have authored at least one GitHub Actions configuration commit. Travis CI follows at 0.40, while Cirrus CI is lowest at 0.14 (Fig.~\ref{fig:rq2_engagement}, Table~\ref{tab:engagement_overall}). The Kruskal--Wallis test across services is significant ($H = 2{,}809.09$, $p < 0.001$) but the effect is small ($\eta^2 = 0.018$). Service choice explains little of the variance. The wide interquartile ranges across all services are the more informative result: in many repositories, CI configuration is handled either by a single contributor or by the whole team, regardless of service. Per-language tests are significant for all seven languages but uniformly small or negligible ($\eta^2$ 0.006--0.038). Removing bot authors leaves the picture unchanged, since bots are 6.5\% of CI authors on GitHub Actions and below 2\% on every other service. The human-only median rises only for Travis CI (0.40 to 0.50) and is otherwise identical to two decimal places. Team-level engagement is therefore not an artifact of automation. Developer engagement in CI is a high-variance, service-independent phenomenon, driven by team-level practice rather than by the CI service.

\begin{table}[t]
\centering
\caption{Developer engagement (author-to-contributor ratio) across CI services.}
\label{tab:engagement_overall}
\begin{tabular}{p{3.5cm}rrrrr}
\toprule
\textbf{CI Service} & \textbf{Repos} & \textbf{Median} & \textbf{Mean} & \textbf{p75} & \textbf{p90} \\
\midrule
GitHub Actions      & \num{89024} & 0.500 & 0.531 & 1.000 & 1.000 \\
Travis CI           & \num{50621} & 0.400 & 0.499 & 1.000 & 1.000 \\
CircleCI            & \num{7134}  & 0.333 & 0.439 & 0.667 & 1.000 \\
GitLab CI           & \num{3991}  & 0.269 & 0.397 & 0.500 & 1.000 \\
AppVeyor            & \num{4851}  & 0.250 & 0.385 & 0.500 & 1.000 \\
Azure Pipelines     & \num{1466}  & 0.250 & 0.369 & 0.500 & 1.000 \\
Bitbucket Pipelines & 229         & 0.333 & 0.392 & 0.500 & 1.000 \\
Cirrus CI           & 379         & 0.143 & 0.267 & 0.333 & 1.000 \\
\bottomrule
\end{tabular}
\end{table}

\smallskip\subsubsection{\textbf{RQ2.4: Temporal Abandonment Patterns}}~

\smallskip\noindent\textbf{Approach.}
To characterize \emph{when} abandonment occurs, we mapped each abandoned repository-service pair to the year of its last CI-related commit for that service. Considering that abandonment is defined by the absence of a configuration file in the current snapshot rather than by an explicit removal timestamp, the year of the last CI commit is the best available proxy for when the service was dropped. We aggregated abandonment events by year and service, reporting raw counts and within-year shares, overall and per language, and tested whether abandonment year differs across services with Kruskal--Wallis and $\eta^2$, applying Bonferroni correction across the seven per-language tests (adjusted $\alpha \approx 0.0071$).

\smallskip\noindent\textbf{Findings.}
Abandonment timing separates into three phases. Through 2018, Travis CI accounts for the large majority of abandonment events, from 100\% in the earliest years to about 40--55\% by 2017 and 2018, with CircleCI making up most of the remainder. From 2019 to 2021, abandonment surges as Travis CI withdraws free access for open-source projects: annual Travis CI abandonment events grow from \num{1625} in 2019 to \num{5306} in 2021, when Travis CI alone accounts for 71.7\% of all abandonment, the single most concentrated abandonment signal in the data. From 2022 onward, Travis CI abandonment declines as its user base shrinks, while GitHub Actions' share of abandonment events rises from 22.9\% in 2022 to 51.2\% in 2024. This rise is a denominator effect rather than a sign of platform failure. The annual abandonment pool contracts from \num{4580} events in 2022 to \num{2812} in 2024 as Travis CI abandonment collapses from \num{2466} to 805 events, while GitHub Actions contributes a steady \num{1000} to \num{1650} removals a year from its very large adopter base. Because GitHub Actions retains configurations at the highest rate of any service (RQ2.5), these removals are a small fraction of its adopters. RQ2.6 shows they are largely true exits from the observed CI set rather than migrations to another service. Abandonment year differs significantly across services overall ($H = 5{,}669.64$, $p \approx 0$, $\eta^2 = 0.180$, large) and in every language after correction ($\eta^2$ from 0.112 in Ruby to 0.288 in Go). The Travis CI collapse is visible in all seven languages, most heavily in Python (\num{1743} Travis CI abandonments in 2021), and by 2024 GitHub Actions accounts for the plurality of abandonment events in every language, from 42.4\% in Rust to 58.5\% in Java.

\smallskip\subsubsection{\textbf{RQ2.5: Survival Analysis of Time-to-Abandonment}}~

\smallskip\noindent\textbf{Approach.}
To model how long services are retained before abandonment, we applied Kaplan--Meier survival analysis~\cite{kaplanmeier1958}. For each repository-service pair, the event of interest is abandonment. For abandoned pairs, duration is the number of days from the first to the last CI commit. For retained pairs, it is the number of days from the first CI commit to the study reference date. We excluded zero-duration pairs. We fit one survival curve per service using the \texttt{lifelines} library~\cite{lifelines}. We then tested for differences across services with the multivariate log-rank test~\cite{peto1972}, applying Bonferroni correction across the seven per-language tests ($\alpha \approx 0.0071$).
The log-rank test establishes that survival differs across services but cannot say whether that difference survives controlling for repository characteristics. We complemented it with a Cox proportional-hazards model~\cite{cox1972} over the \num{161309} repository-service pairs with a nonzero retention span (the same nonzero-duration subset used for the survival curves). The model regresses the abandonment hazard on a multi-CI indicator; log-transformed repository age, stars, and contributor count; and one-hot service and language terms. It uses a small ridge penalty for numerical stability and reports hazard ratios with 95\% confidence intervals and Harrell's concordance. We omit configuration size as a covariate: it is measured on the current snapshot, and an abandoned service therefore has no file left to count. The covariate is thus both missing for essentially every event and measured after the outcome. We analyzed the proportional-hazards assumption with the scaled Schoenfeld residual test~\cite{grambsch1994}. Given that assumption is known to fail under large samples and no service reaches median survival within the window, we also report the restricted mean survival time (RMST)~\cite{royston2013}. RMST is the area under each service's survival curve up to a shared horizon, which we set to the shortest per-service maximum follow-up (\num{2302} days). It summarizes retention without assuming proportional hazards or a reachable median.

\begin{figure}[htbp]
\centering
\includegraphics[width=0.9\linewidth]{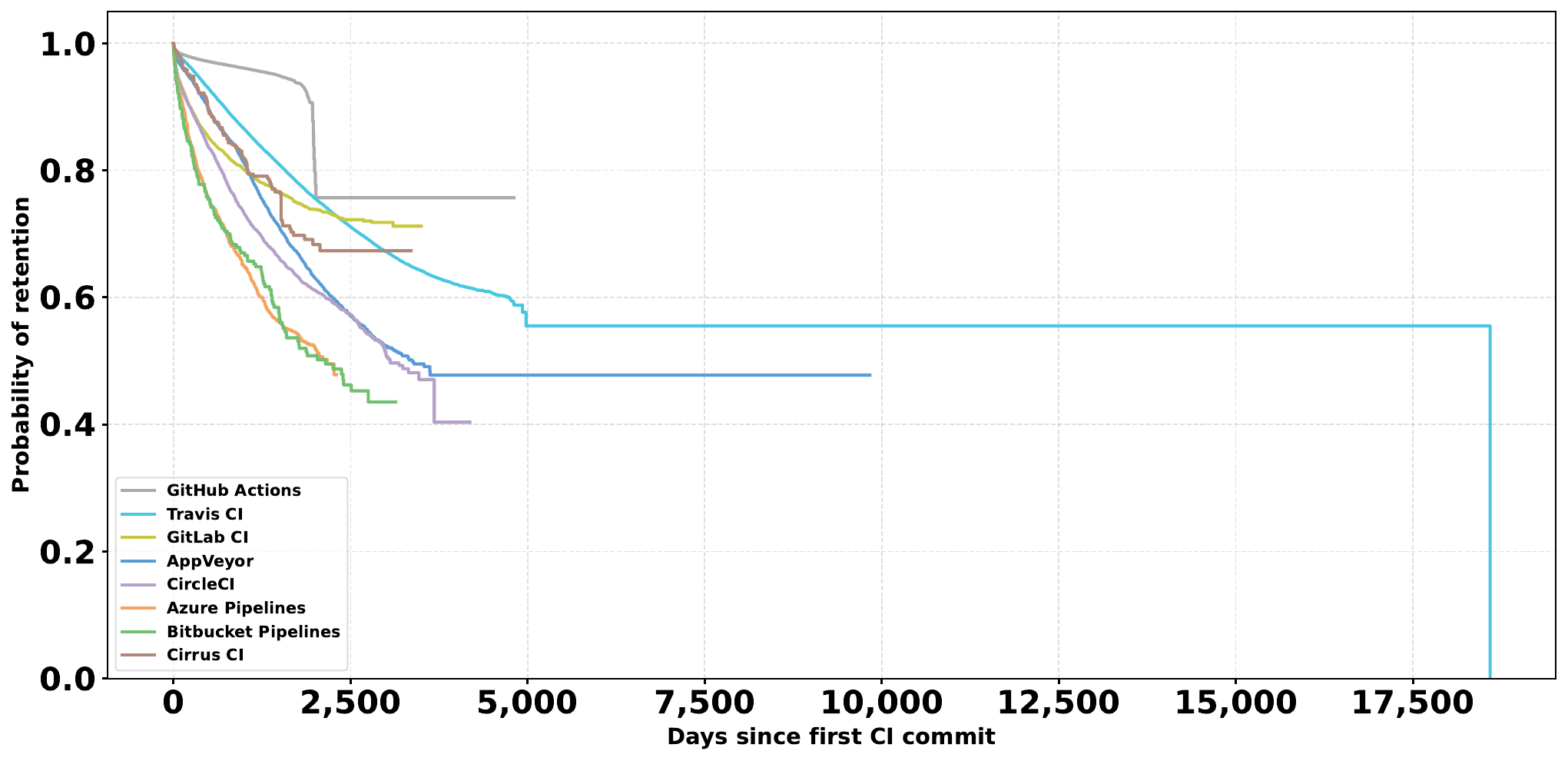}
\Description{Kaplan--Meier survival curves showing the probability that repositories continue using each CI service over time before abandonment, across all programming languages. The horizontal axis represents time since CI adoption and the vertical axis represents the estimated probability of continued use. Curves that remain higher indicate longer-lasting adoption and greater retention, while steeper declines indicate more frequent abandonment.}
\caption{Kaplan--Meier survival curves for CI service abandonment across all languages. Higher curves indicate greater retention. Curves for low-frequency services (Azure Pipelines, Bitbucket Pipelines, and Cirrus CI) are stepped and end earlier because few repositories remain at risk at longer durations. Flat terminal segments reflect sparse observations rather than stable adoption.}
\label{fig:rq2_survival}
\end{figure}

\smallskip\noindent\textbf{Findings.}
The log-rank test rejects equal survival across services overall (test statistic $= 9{,}353.57$, $p \approx 0$) and in every language after correction (all $p < 10^{-50}$; Fig.~\ref{fig:rq2_survival}). No service reaches a median survival time within the study window, meaning that in no service has the majority of adopters removed the configuration. The curves nonetheless diverge widely in how fast they decline. The event rates reported here are computed on the survival subset, which excludes zero-duration pairs. They therefore run below the abandonment rates in Table~\ref{tab:abandon}. The difference is most noticeable for GitHub Actions (4.1\% vs. 6.6\%), which, as the most recently adopted service, contributes the most zero-duration pairs. GitHub Actions is the most durable, with only 4.1\% of its pairs abandoned and a curve near 1.0 throughout, followed by GitLab CI (23.6\%) and Cirrus CI (25.2\%), whose tight platform integration limits substitution. Travis CI (31.5\%), CircleCI (37.6\%), AppVeyor (42.0\%), Azure Pipelines (44.6\%), and Bitbucket Pipelines (48.8\%) decline faster, with Bitbucket Pipelines approaching but not crossing the 50\% threshold. Per pair, the least-integrated services decline fastest. However, Travis CI and CircleCI account for the bulk of abandonment in absolute terms, since they combine an elevated rate with by far the largest adopter bases. The lifecycle narrative therefore centers on them rather than on the higher-rate but small-population services. This pattern is consistent across languages, with AppVeyor abandonment highest in the systems languages (C, C++, Rust). A median survival time can be defined only for the few language-service cells whose event rate exceeds 50\%. All of them fall among the less-integrated services: Azure Pipelines in Rust (85 of 133 pairs, 63.9\%); Bitbucket Pipelines in C++ (30 of 52, 57.7\%) and Python (43 of 80, 53.8\%); and AppVeyor in Python (559 of \num{1052}, 53.1\%) and Rust (320 of 630, 50.8\%). The two AppVeyor cases echo its concentration in the Windows-oriented languages observed throughout RQ1.

The Cox model confirms that these service differences persist after controlling for repository age, popularity, and team size (concordance 0.774 on \num{161309} pairs, \num{27315} events; Table~\ref{tab:cox}). Relative to AppVeyor, GitHub Actions has less than half the abandonment hazard (HR 0.40). The highest per-pair hazards fall on the small-population services Azure Pipelines (HR 2.17) and Bitbucket Pipelines (HR 2.03). CircleCI (HR 1.66) and Travis CI (HR 1.52) are also elevated, reproducing the survival-curve ordering under adjustment. Travis CI and CircleCI nonetheless matter most for the overall lifecycle, since their elevated hazard acts on far larger adopter bases. Two covariate effects are noteworthy. Older repositories are far more durable (HR 0.57 per unit of log age), making age the strongest continuous predictor. Stars and contributor count have only marginal effects (HR 1.08 and 1.04). The multi-CI indicator is associated with the largest hazard ratio in the model (HR 6.32). A service is much more likely to be dropped when the repository runs others alongside it. This is the per-service imprint of the migration dynamic, since the service being left behind co-exists with its replacement for a time. The proportional-hazards assumption is rejected for most covariates, but this is expected at this sample size. As discussed in Section~\ref{sec:threats}, we therefore consider the hazard ratios as average effects over the window and report RMST alongside. RMST corroborates these findings without that assumption. Within the \num{2302}-day horizon, GitHub Actions retains \num{2143} days (93.1\% of the horizon), ahead of Travis CI (\num{1961}). Azure Pipelines (\num{1496}) and CircleCI (\num{1686}) are retained least, matching the survival-curve and Cox orderings.

\begin{table}[htbp]
\centering
\caption{Cox proportional-hazards model of time-to-abandonment (\num{161309} pairs, \num{27315} events, concordance 0.774). Hazard ratios are relative to AppVeyor (service) and C (language); HR${}>1$ indicates a higher abandonment hazard. Selected covariates.}
\label{tab:cox}
\begin{tabular}{p{5cm}rr}
\toprule
\textbf{Covariate} & \textbf{HR} & \textbf{95\% CI} \\
\midrule
Multi-CI repository        & 6.32 & [6.13, 6.51] \\
Azure Pipelines (service)  & 2.17 & [1.99, 2.37] \\
Bitbucket Pipelines (service) & 2.03 & [1.70, 2.43] \\
CircleCI (service)         & 1.66 & [1.58, 1.75] \\
Travis CI (service)        & 1.52 & [1.46, 1.58] \\
GitHub Actions (service)   & 0.40 & [0.38, 0.42] \\
Log repository age         & 0.57 & [0.56, 0.58] \\
Log stars                  & 1.08 & [1.07, 1.09] \\
Log contributors           & 1.04 & [1.03, 1.05] \\
\bottomrule
\end{tabular}
\end{table}

\smallskip\subsubsection{\textbf{RQ2.6: Migration versus True Abandonment}}~

\smallskip\noindent\textbf{Approach.}
The abandonment rates in RQ2.1 count any service whose configuration disappears, which conflates two distinct events: migration to another CI service and true exit from the observed CI service set. We therefore classify each abandoned repository-service pair as a \emph{migration} or a \emph{true exit}. A dropped service is considered a migration when the same repository brings another service online around the time the dropped one becomes inactive: specifically, the replacement's first CI commit must fall within 180 days before to 365 days after the last commit to the abandoned service, and the replacement must still be present in the snapshot. All other abandonments are treated as true exits. We report migration shares by source service, construct a flow matrix of conditional destination probabilities, and cross-tabulate GitHub Actions abandonment outcomes by year to test the RQ2.4 interpretation of its rising abandonment share.

\smallskip\noindent\textbf{Findings.}
Half of all abandonments are migrations rather than departures from CI: of the \num{31485} abandoned repository-service pairs, almost exactly half (50.5\%) are replacements and the rest (49.5\%) are exits from the observed CI service set, with a highly service-dependent split (Table~\ref{tab:migration}). Travis CI abandonments are predominantly migrations (66.3\%), consistent with the mass move off the platform after its pricing change, and Azure Pipelines (63.9\%) and AppVeyor (57.0\%) are also migration-heavy. The one service abandoned almost entirely through true exits is GitHub Actions (97.2\% true exits), since a project dropping GitHub Actions is usually leaving CI rather than moving on, and its rare migrations scatter across CircleCI, Travis CI, and GitLab CI rather than converging anywhere. Every other service migrates predominantly \emph{to} GitHub Actions: 93\% of Travis CI migrations, 96\% of Azure Pipelines migrations, and 81--89\% of CircleCI, AppVeyor, and GitLab CI migrations land on it. This decomposition refines RQ2.1, since a large part of what the raw rates call abandonment is the visible trace of consolidation onto a single platform. It also reframes the rise of GitHub Actions in the RQ2.4 abandonment timeline. GitHub Actions removals in 2022--2024 are 98.5\% true exits (\num{4076} of \num{4137}), and its rising share of abandonment events therefore reflects repositories leaving the observed CI set from a very large adopter base rather than any weakness of GitHub Actions itself.

\begin{table}[htbp]
\centering
\caption{Migration versus true exit among abandoned repository-service pairs. Migration\% is the share of that service's abandonments that were replaced by another service still present in the snapshot.}
\label{tab:migration}
\begin{tabular}{p{3.5cm}rrr}
\toprule
\textbf{Service (source)} & \textbf{Abandoned} & \textbf{Migration \%} & \textbf{True-exit \%} \\
\midrule
GitHub Actions      & \num{6049}  & 2.8  & 97.2 \\
Travis CI           & \num{17893} & 66.3 & 33.7 \\
CircleCI            & \num{3242}  & 46.8 & 53.2 \\
AppVeyor            & \num{2204}  & 57.0 & 43.0 \\
GitLab CI           & \num{1144}  & 46.8 & 53.2 \\
Azure Pipelines     & 710         & 63.9 & 36.1 \\
Bitbucket Pipelines & 139         & 48.9 & 51.1 \\
Cirrus CI           & 104         & 31.7 & 68.3 \\
\midrule
\textbf{Overall}    & \num{31485} & 50.5 & 49.5 \\
\bottomrule
\end{tabular}
\end{table}

\begin{rqsummary}{RQ2 Summary}
The CI service lifecycle is dominated by the collapse of Travis CI after its 2020 move to a paid model, the single largest abandonment signal in the data and visible in every language. About 19\% of repositories switch services over time, almost always toward GitHub Actions, which is uniquely durable (4.1\% abandoned, and less than half the abandonment hazard of any other service once age and popularity are controlled) while less integrated services such as AppVeyor and Bitbucket Pipelines are dropped far more often. Half of all abandonments are in fact migrations rather than departures from CI, and 93\% of Travis CI migrations land on GitHub Actions. Much of the measured abandonment is therefore consolidation onto one platform. Developer engagement in CI upkeep is high-variance and driven by team practice, not by the service.
\end{rqsummary}

\subsection{RQ3: Why Developers Adopt, Switch, and Abandon CI Services}
\label{subsec:rq3}

\noindent\textbf{Motivation.}
RQ1 and RQ2 establish what developers do with CI services and when, but not why. The reasons behind CI decisions have so far been reported mainly based on interviews with small numbers of practitioners~\cite{rostami2023usage}. Commit messages offer a complementary, large-scale record of the rationale developers attach to their own configuration changes, and this question mines that record to ask why services are adopted, co-adopted, migrated, and abandoned, and how far commit messages can answer that question at all.

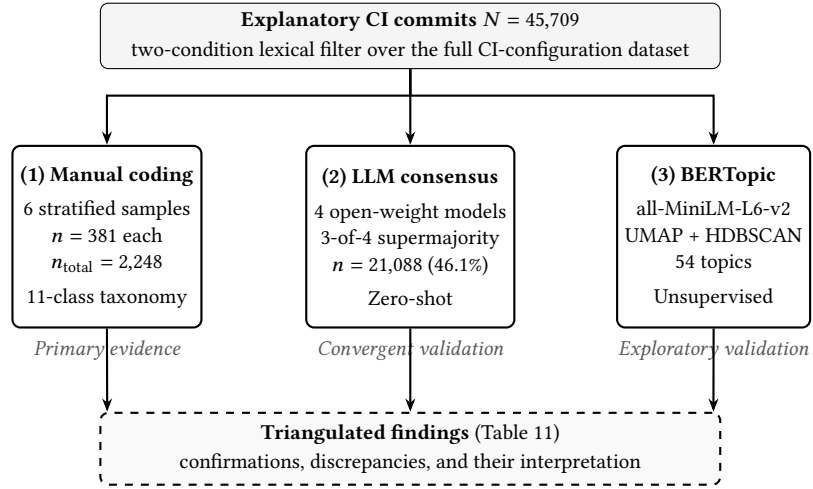
\begin{figure}[ht]
  \centering
  \begin{tikzpicture}[
    font=\small,
    wbox/.style={draw, rounded corners=3pt, minimum width=8.2cm, minimum height=0.85cm, align=center, fill=gray!8},
    mbox/.style={draw, rounded corners=3pt, thick, minimum width=2.5cm, minimum height=2.4cm, align=center, fill=white, font=\small},
    obox/.style={draw, rounded corners=3pt, dashed, thick, minimum width=8.2cm, minimum height=0.95cm, align=center, fill=gray!5},
    arr/.style={-{Stealth[length=5pt]}, thick},
    arrdash/.style={-{Stealth[length=5pt]}, thick, dashed},
    lbl/.style={font=\small\itshape, text=gray!70!black}
  ]
  \node[wbox] (set) at (0,0)
    {\textbf{Explanatory CI commits}~~$N = \num{45709}$\\[1pt]
     \small two-condition lexical filter over the full CI-configuration dataset};
  \node[mbox, below left=1.0cm and 2.749cm of set.south] (manual)
    {\textbf{(1) Manual coding}\\[3pt]
     6 stratified samples\\
     $n=381$ each\\
     $n_{\text{total}}=\num{2248}$\\[3pt]
     11-class taxonomy};
  \node[mbox, below=1.0cm of set.south] (llm)
    {\textbf{(2) LLM consensus}\\[3pt]
     4 open-weight models\\
     3-of-4 supermajority\\
     $n=\num{21088}$ (46.1\%)\\[3pt]
     Zero-shot};
  \node[mbox, below right=1.0cm and 2.722cm of set.south] (bert)
    {\textbf{(3) BERTopic}\\[3pt]
     all-MiniLM-L6-v2\\
     UMAP + HDBSCAN\\
     54 topics\\[3pt]
     Unsupervised};
  \draw[arr] (set.south) -- ++(0,-0.35) -| (manual.north);
  \draw[arr] (set.south) -- (llm.north);
  \draw[arr] (set.south) -- ++(0,-0.35) -| (bert.north);
  \node[lbl, below=1pt of manual.south] {Primary evidence};
  \node[lbl, below=1pt of llm.south]    {Convergent validation};
  \node[lbl, below=1pt of bert.south]   {Exploratory validation};
  \node[obox, below=1.1cm of llm.south] (findings)
    {\textbf{Triangulated findings} (Table~\ref{tab:rq3-synthesis})\\[1pt] \small confirmations, discrepancies, and their interpretation};
    \coordinate (findings-left) at ($(findings.north)+(-4.01cm,0)$);
  \coordinate (findings-right) at ($(findings.north)+(4.01cm,0)$);
  \draw[arr] (manual.south) -- (findings-left);
  \draw[arr] (llm.south) -- (findings.north);
  \draw[arr] (bert.south) -- (findings-right);  \end{tikzpicture}
  \Description{Diagram showing the RQ3 analysis approach. The same set of CI-configuration commits is analyzed using three methods: manual coding, LLM consensus classification, and BERTopic modeling. All methods use a shared 11-class taxonomy consisting of four strategic categories (A1--A4) and six maintenance-context categories (M1--M6). Manual coding provides the primary category proportions, while LLM consensus and BERTopic provide validation and exploratory analysis of the rationale categories.}
  \caption{RQ3 approach: three methods applied to the same candidate set using one 11-class taxonomy (A1--A4 strategic, M1--M6 maintenance-context). Manual coding provides the primary evidence for proportions; LLM consensus and BERTopic provide convergent and exploratory validation.}
  \label{fig:rq3-overview}
\end{figure}

\smallskip\noindent\textbf{Approach.}
\label{subsec:rq3-method}
We analyzed the natural-language rationale developers embed in commit messages that touch CI configuration files. We used a triangulated design over a shared candidate set of \num{45709} explanatory commits (Fig.~\ref{fig:rq3-overview}). From the full dataset of CI-configuration commits, we retained only explanatory ones through a two-condition lexical filter. The filter requires both a causal phrase (such as \texttt{because}, \texttt{since}, \texttt{due to}, \texttt{no longer}, \texttt{deprecated}, or \texttt{instead of}) and a reference to a CI service or concept (such as \texttt{travis}, \texttt{github actions}, \texttt{pipeline}, or \texttt{workflow}). This procedure follows the intent-extraction approach of Widder et al.~\cite{widder2018} and Zampetti et al.~\cite{zampetti2017}. The filter retains only \num{45709} of the \num{2450710} CI-configuration commits (1.9\%). This full dataset exceeds the \num{2423030}-commit maintenance sample of RQ1.4, which excludes 83 repositories with corrupted commit counts. The filter is intentionally conservative: it keeps the commits most likely to state a reason and discards the rest. The strategic-decision rates we report for this explanatory subset are upper bounds on the rate across all CI activity.

We employed manual qualitative coding, LLM consensus classification, and topic modeling in a three-method design. The primary analysis is manual qualitative coding by the first co-author against an 11-class taxonomy. The taxonomy was developed from both the interview-based categories~\cite{rostami2023usage} and patterns that emerged during our analysis of the commits~\cite{braun2006,strauss1998}. It comprises four strategic categories that capture explicit CI decisions (A1--A4), six maintenance-context categories that capture the operational circumstances surrounding them (M1--M6), and a residual class (IR: irrelevant). The residual class covers commits that survive the lexical filter but provide no CI-relevant reason, as detailed in Table~\ref{tab:rq3-taxonomy}. Commits that are labeled IR are excluded from the reported distribution, and the proportions below therefore cover only the ten substantive categories. Strategic decisions are rare in configuration-commit messages, since choices such as co-adoption are usually settled in design discussions rather than announced in a single commit, which is likely why prior work studied them through developer interviews~\cite{rostami2023usage}. A single representative sample of about 381 commits would contain only about two co-adoption commits and about ten each of initial adoption and abandonment, too few to characterize these rare categories on their own. We therefore drew six independent stratified random samples of 381 commits each ($95\%$ confidence level, $\pm 5\%$ margin of error).
After removing overlaps, we combined the samples into \num{2248} unique commits. The category distribution was stable across the six draws, with the per-sample strategic share ranging over 7.6--13.4\%. The resulting estimate is therefore not driven by any single sample. As convergent validation, we classified all \num{45709} candidates with four open-weight LLMs (\texttt{llama3.1-8b}, \texttt{gemma3-12b}, \texttt{mistral-7b}, \texttt{phi4-14b}). A supermajority filter retains only commits for which at least three of four models agree ($n = \num{21088}$, 46.1\% of candidates)~\cite{abrokwah2025compliant}. This design follows evidence that zero-shot open-weight models underperform fine-tuned classifiers on specialized tasks~\cite{bucher2024}. As exploratory validation, we fit BERTopic~\cite{grootendorst2022} on all \num{45709} candidates. We treat the qualitative coding as primary evidence for proportions and the two automated methods as directional and exploratory validation~\cite{creswell2018}, reporting confirmations, discrepancies, and their interpretation.

Two further checks probe whether the reason evidence is distorted by automation and whether it connects to observed behavior. To test the automation concern, we classified all \num{2450710} CI-configuration commits with a keyword scheme aligned to the taxonomy, identified bot authors with the rule from RQ1.4, and recomputed the category shares with bot commits removed. Any category inflated by automated commits would then show a large shift. To test whether stated reasons track real outcomes, we linked the reason evidence back to RQ2. We flagged every CI-configuration commit whose message indicates an exit-intent signal through migration or service-removal vocabulary. We then measured how often the corresponding repository-service pair is actually abandoned and compared that conditional rate with the population base rate using a chi-squared test and the ratio between the two rates.

\begin{table}[t]
\centering
\caption{Commit-message label taxonomy. Strategic categories (A1--A4) are direct answers to RQ3. Maintenance-context categories (M1--M6) capture the operational circumstances around strategic decisions. Percentages cover the ten substantive categories (IR excluded), rounded to one decimal and therefore not necessarily summing to exactly 100.}
\label{tab:rq3-taxonomy}
\small
\begin{tabular}{lllr}
\toprule
\textbf{Label} & \textbf{Type} & \textbf{Description} & \textbf{\%} \\
\midrule
\multicolumn{4}{l}{\textit{\textbf{Strategic categories}}} \\[2pt]
A1 & Strategic & Initial adoption of a CI service & 3.0 \\
A2 & Strategic & Migration between services, both identifiable & 4.8 \\
A3 & Strategic & Co-adoption of a second service alongside one in use & 0.4 \\
A4 & Strategic & Abandonment of a CI service & 2.8 \\
\midrule
\multicolumn{4}{l}{\textit{\textbf{Maintenance-context categories}}} \\[2pt]
M1 & Maintenance & Deprecation or EOL reaction & 14.5 \\
M2 & Maintenance & Reliability fix (flaky, broken, transient) & 21.4 \\
M3 & Maintenance & Performance or speed optimization & 8.1 \\
M4 & Maintenance & Resource constraint (memory, timeout, free-tier) & 6.6 \\
M5 & Maintenance & Platform or OS constraint & 4.7 \\
M6 & Maintenance & Configuration or dependency fix, no detectable reason & 33.9 \\
\bottomrule
\end{tabular}
\end{table}

\smallskip
\smallskip\noindent\textbf{Findings.}

\smallskip\noindent\textbf{CI configuration activity is predominantly reactive rather than strategic.}
Only about 11\% of explanatory commits are related to a strategic CI decision (A1--A4: initial adoption 3.0\%, migration 4.8\%, co-adoption 0.4\%, abandonment 2.8\%). The remaining 89\% are maintenance-context commits that reveal why CI configuration requires ongoing effort but record no deliberate choice of service, and this split holds across all six samples and regardless of language or service. Reactive maintenance, the deprecation reactions (M1) and reliability fixes (M2) taken together, is the single largest group at 35.9\%. The operational burden captured by these categories is a structural force that, accumulated over time, motivates the strategic migrations and abandonments observed in RQ2.

\smallskip\noindent\textbf{Reliability of the coding.}
A stratified subset of 381 coded commits (16.9\%), spanning all six samples and all seven languages, was independently labeled by the second co-author. For the strategic-versus-maintenance distinction on which RQ3's conclusions depend, the two codings agreed on 97.4\% of commits, indicating that the reactive-maintenance majority is not sensitive to the identity of the coder. On the full 11-class taxonomy, observed agreement was 87.9\% and Cohen's $\kappa = 0.85$ (95\% CI [0.81, 0.89])~\cite{landis1977}, in the almost-perfect range and well above the $\kappa \geq 0.70$ threshold commonly used for qualitative classification. Agreement was high across the well-populated maintenance categories, reaching 92\% for reliability fixes (M2) and 90\% for generic config-maintenance (M6), with deprecation reactions (M1) the lowest among the common categories at 77\%. The strategic categories and the platform-constraint category (M5) are individually small, holding no more than 21 commits each in the subset. Their per-category rates are unstable, and we do not read into them. The two coders disagreed on 46 of the 381 commits (12.1\%), of which only 10 crossed the strategic-versus-maintenance boundary. This is why the higher-level split shows 97.4\% agreement while the full taxonomy shows 87.9\%. About half of the disagreements (25 of 46) turn on the M6 boundary, namely whether a commit states a detectable reason such as a deprecation reaction (M1) or a reliability fix (M2) or instead records only a generic configuration edit (M6). Most of the remainder fall between adjacent maintenance categories such as deprecation vs. reliability (M1 vs. M2). Both patterns reflect real ambiguity in developer intent rather than coder error, since one short commit message can be read as either reason-bearing or routine. We resolved every disagreement by discussion between the two coders, returning to the taxonomy definitions until each commit reached a consensus label and clarifying the codebook wording where a boundary proved systematically ambiguous. Reconciliation moved a handful of commits between maintenance categories but left the reactive-maintenance majority and the strategic-versus-maintenance split unchanged.

\smallskip\noindent\textbf{Strategic decisions: migration, adoption, abandonment, co-adoption.}
Migration (A2, 4.8\%) is the most common strategic category, and among migration commits with an extractable direction, about a third correspond to a Travis CI to GitHub Actions transition, confirming at scale the dominant pattern from RQ2.2 and from interviews~\cite{rostami2023usage}. Commit messages rarely explain why the destination was chosen. Where they do, three drivers recur: native platform integration (crediting GitHub Actions' embedding in GitHub), reliability failure of the source service (Travis CI ``not working anymore''), and speed, which is bidirectional, with a minority of projects moving to CircleCI because GitHub Actions was too slow for compute-intensive work. Initial adoption (A1, 3.0\%) is dominated historically by Travis CI and, after 2019, primarily by GitHub Actions. Abandonment (A4, 2.8\%) partitions into five subtypes, of which the most common is post-migration cleanup, where the configuration is removed after the migration decision was already made in an earlier commit. This has a methodological consequence: studies that date abandonment from a configuration-file deletion systematically attribute it to the wrong commit and understate the true interval between decision and cleanup.

Co-adoption rationale (A3) is the rarest strategic category at 0.4\%, and its scarcity is itself a finding. Interviews report co-usage as common and multi-motivated~\cite{rostami2023usage}, yet it is almost invisible in commit messages, since adding a second service is an architectural decision made in design discussions rather than announced in a single commit. The few A3 commits we find describe staged migration, functional specialization, a performance escape hatch (adding CircleCI because GitHub Actions was slow), and exploratory trials.

\smallskip\noindent\textbf{Maintenance context: what drives the ongoing effort.}
The stated-reason maintenance categories expose the forces that accumulate pressure toward strategic decisions. Deprecation and EOL reactions (M1, 14.5\%) are the largest explicitly stated reason and are reactive by definition, driven by provider-imposed runner-image changes, language-runtime EOLs, and, in a pattern not present in prior interview work, the abandonment of unmaintained third-party Actions from the GitHub Actions Marketplace. Reliability fixes (M2, 21.4\%) are the most prevalent category overall and cluster around networking instability, unannounced environment drift, upstream dependency breakage, flakiness, and provider bugs, with several commits naming a reliability failure in one service as the trigger for considering migration to another. Performance work (M3, 8.1\%) is mostly caching, parallelization, and matrix reduction within an existing service rather than a reason to switch. Resource constraints (M4, 6.6\%) name symptoms, not pricing: the Travis CI 50-minute build limit, memory and out-of-memory errors, and free-tier storage and concurrency caps drive concrete restructuring, but developers write about minutes and memory, never about billing. Cost-driven decisions are systematically under-detected by commit-message analysis. Platform and OS constraints (M5, 4.7\%) surface architecture gaps, missing hardware such as GPUs on runners, stale system libraries specific to the Travis CI platform in C and C++ projects, and an emerging concern around supply-chain security, reflected in the pinning of third-party Actions to immutable commit SHAs. Operational configuration activity with no detectable reason (M6, 33.9\%) is the largest single category and quantifies how much CI activity offers no information about service decisions, a direct reflection of the configuration complexity documented in RQ1.5.

\smallskip\noindent\textbf{Triangulation with LLM consensus and topic modeling.}
The two automated methods corroborate the central conclusion while clarifying its limits (Table~\ref{tab:rq3-synthesis}). The LLM consensus distribution matches the qualitative coding closely on the diagnostic categories, confirming that co-adoption is rarely expressed in commit messages (0.6\% vs. 0.4\%) and that initial adoption accounts for about 3\% of cases. It also preserves the ordinal ranking of maintenance categories, with GitHub Actions commits dominated by deprecation reactions and Travis CI commits by reliability fixes. It over-classifies migration (11.3\% vs. 4.8\%), as models flag migration vocabulary without verifying the source-to-destination structure the manual criterion requires. The LLM migration figure is thus an upper bound. BERTopic reduces to 54 topics, of which only three map to strategic decisions while the rest are operational. Its single largest topic ($n = \num{19807}$) is a semantic catch-all of generic CI vocabulary corresponding to M6. The unsupervised clustering cleanly separates operational patterns such as the Node.js deprecation cascade and the AppVeyor Windows platform gap, but it cannot separate strategic intent from routine maintenance. This contrast provides the strongest evidence that commit messages are effective for studying CI maintenance but poor at revealing the strategic reasons behind CI service choices. Those reasons likely reside in pull-request descriptions/discussions or issue threads rather than commits.

\begin{table}[htbp]
\centering
\caption{Triangulation confidence across the qualitative coding ($n = \num{2248}$), LLM consensus ($n = \num{21088}$), and BERTopic ($n = \num{45709}$). \checkmark\checkmark\checkmark~= all three agree; \checkmark\checkmark~= two agree; \checkmark~= coding only.}
\label{tab:rq3-synthesis}
\small
\begin{tabular}{lc}
\toprule
\textbf{Finding} & \textbf{Support} \\
\midrule
Travis CI to GitHub Actions is the dominant migration      & \checkmark\checkmark\checkmark \\
Deprecation/EOL is the largest stated-reason category      & \checkmark\checkmark\checkmark \\
Co-adoption motivation is rare ($\approx$0.4\%)            & \checkmark\checkmark\checkmark \\
Initial adoption $\approx$3\%                              & \checkmark\checkmark\checkmark \\
GitHub Actions commits dominated by deprecation reactions  & \checkmark\checkmark\checkmark \\
Travis CI commits dominated by reliability fixes           & \checkmark\checkmark\checkmark \\
Free-tier constraints named as symptoms, not pricing       & \checkmark\checkmark\checkmark \\
M6 (no reason) is the largest single category ($\approx$34\%) & \checkmark\checkmark\checkmark \\
Supply-chain SHA pinning as an emergent M5 sub-pattern     & \checkmark\checkmark \\
GitHub Actions too slow, CircleCI added as supplement      & \checkmark\checkmark \\
Strategic decisions $\approx$11\% of explanatory commits   & \checkmark \\
Abandonment subtypes (post-migration cleanup, etc.)        & \checkmark \\
\bottomrule
\end{tabular}
\end{table}

\smallskip\noindent\textbf{Robustness to automation and link to observed outcomes.}
Two additional analyses strengthen our interpretation that commit-message rationales provide meaningful, though incomplete, evidence of CI decision-making. First, automated commits do not distort the reason mix. Bots author 6.4\% of the full \num{2450710} CI-configuration commits (\num{157534}), the same low share found in the RQ1.4 maintenance subset (6.5\%), concentrated in GitHub Actions (10.1\%) and negligible elsewhere. Recomputing the category shares on human-authored commits only leaves the ordering intact: every strategic category moves by less than 0.2 percentage points, and the largest shift anywhere is a 4.9-point drop in routine maintenance updates, the category most naturally produced by automation. The manual distribution, coded on samples that include bot-authored commits, is therefore not an artifact of automation. Second, the reasons developers state track what they actually do. Commits with an exit-intent signal are followed by real abandonment of that repository-service pair 40.4\% of the time, vs. a 20.9\% base rate, a relative rate of about 1.9 ($\chi^2$ significant at $p < 10^{-3}$, though the association is weak, $\phi = 0.086$). Stated migration and removal intent thus has measurable predictive power on the lifecycle outcomes measured in RQ2, which is the strongest available evidence that commit-message rationale, for all its incompleteness, reflects decisions that are implemented rather than merely discussed.

\begin{rqsummary}{RQ3 Summary}
CI configuration activity is mostly reactive: only about 11\% of commits state a reason related to a strategic decision, while reactive maintenance (deprecation reactions and reliability fixes) alone is 35.9\%. The Travis CI to GitHub Actions migration is confirmed at scale and driven by integration, reliability, and free-tier withdrawal, while co-adoption rationale is nearly absent from commit messages as it is likely decided elsewhere. Commit messages are an effective source for CI maintenance patterns but a poor one for strategic CI reasoning.
\end{rqsummary}

\subsection{RQ4: Predicting Multi-CI Adoption and Abandonment}
\label{subsec:rq4}

\noindent\textbf{Motivation.}
The preceding questions describe how multi-CI adoption and abandonment evolve over time and why developers say they occur. This question asks which repositories they happen to, by modeling the two outcomes from repository characteristics. Identifying what distinguishes multi-CI adopters and predicts abandonment clarifies whether these behaviors are project-specific, separates repository-level from platform-level drivers, and reveals potential confounding factors, such as project age, that future studies of CI diversity should control for.

\smallskip\noindent\textbf{Approach.}
We constructed two datasets from the full study data. The repo-level dataset contains \num{134264} repositories, each with its CI configuration, activity metrics, and metadata. The repo-service dataset contains \num{39127} repository-service pairs, where a repository appears once per adopted service, enabling service-level analysis. We defined three binary outcomes. The first is multi-CI adoption at the repo level (positive class \num{25475}, 19.0\%) The second is CI abandonment at the repo level (positive class \num{27162}, 20.2\%). The third is abandonment of a specific service at the repository-service level (positive class \num{31157}, 79.6\% of the \num{39127}-pair modeling dataset). This modeling subset retains only repository-service pairs with complete commit histories and all required feature values. Thus, its positive rate is not directly comparable to the 18.9\% abandonment rate across all \num{166184} pairs in the full population (Table~\ref{tab:abandon}).

We engineered 23 features across five groups: activity (commits, pull requests, issues, contributors, and their monthly frequencies), ratios (PR-to-commit, issue-to-commit, contributor-to-commit), temporal (repository age, days since last push), popularity (stars, watchers, forks, open issues), and metadata (size, topic count, and flags for organization ownership, license, description, and topics). The repo-service model adds one-hot indicators for each service. Before modeling, we dropped near-zero-variance columns and removed features correlated above $|r| > 0.90$, keeping the first of each pair, which eliminated the watcher count (redundant with stars) and, in the repo-service model, the issue count (redundant with open issues) and standardized the remaining features. We compared groups on each feature with Mann--Whitney tests under Bonferroni correction (adjusted $\alpha \approx 0.0022$ for 23 tests), reporting the rank-biserial correlation as effect size. We then fit logistic regression models with \texttt{statsmodels}, reporting odds ratios with 95\% confidence intervals and McFadden's pseudo-$R^2$, and quantified each feature's contribution with SHAP~\cite{lundberg2017shap}. The large positive-class counts keep events per variable well above standard guidelines for all three models~\cite{peduzzi1996}.

Given that pseudo-$R^2$ measures fit rather than out-of-sample performance, we also assessed discrimination directly with 5-fold stratified cross-validation. We report the ROC-AUC, the precision-recall AUC (PR-AUC) relative to the class base rate, and the Brier score for calibration. We checked residual collinearity among the retained features with the variance inflation factor~\cite{belsley1980}, flagging any feature above the conventional thresholds of 5 and 10. To test whether the linear form of logistic regression hides nonlinear structure, we benchmarked each outcome against a gradient-boosting classifier~\cite{friedman2001} under the same cross-validation. Finally, the repo-service abandonment model is fit on a subset enriched in abandonment (79.6\% positive). We therefore refit it on the full detected-pair population: \num{165039} pairs, of which 18.9\% are abandoned. This is the Table~\ref{tab:abandon} population of \num{166184} pairs minus those dropped in feature construction. We compared the coefficient signs to establish which conclusions transfer beyond the enriched subset.

\smallskip\noindent\textbf{Findings.}

Multi-CI repositories ($n = \num{25475}$) differ from single-CI repositories ($n = \num{108789}$) on every feature but one. They are substantially more active and mature, with median commits of 260 vs. 78 ($r = 0.42$, medium), median contributors of 6 vs. 2 ($r = 0.41$, medium), and median age of \num{2462} vs. \num{1561} days ($r = 0.39$, medium), and they attract more community engagement (median stars 50 vs. 7, forks 9 vs. 2). They are also more recently active, with a median of 136 days since last push vs. 441. Multi-CI use is associated with ongoing maintenance rather than dormancy. Whether forking is allowed is the only feature showing no difference ($p = 1.0$).

\begin{figure}[htbp]
\centering
\includegraphics[width=0.7\linewidth]{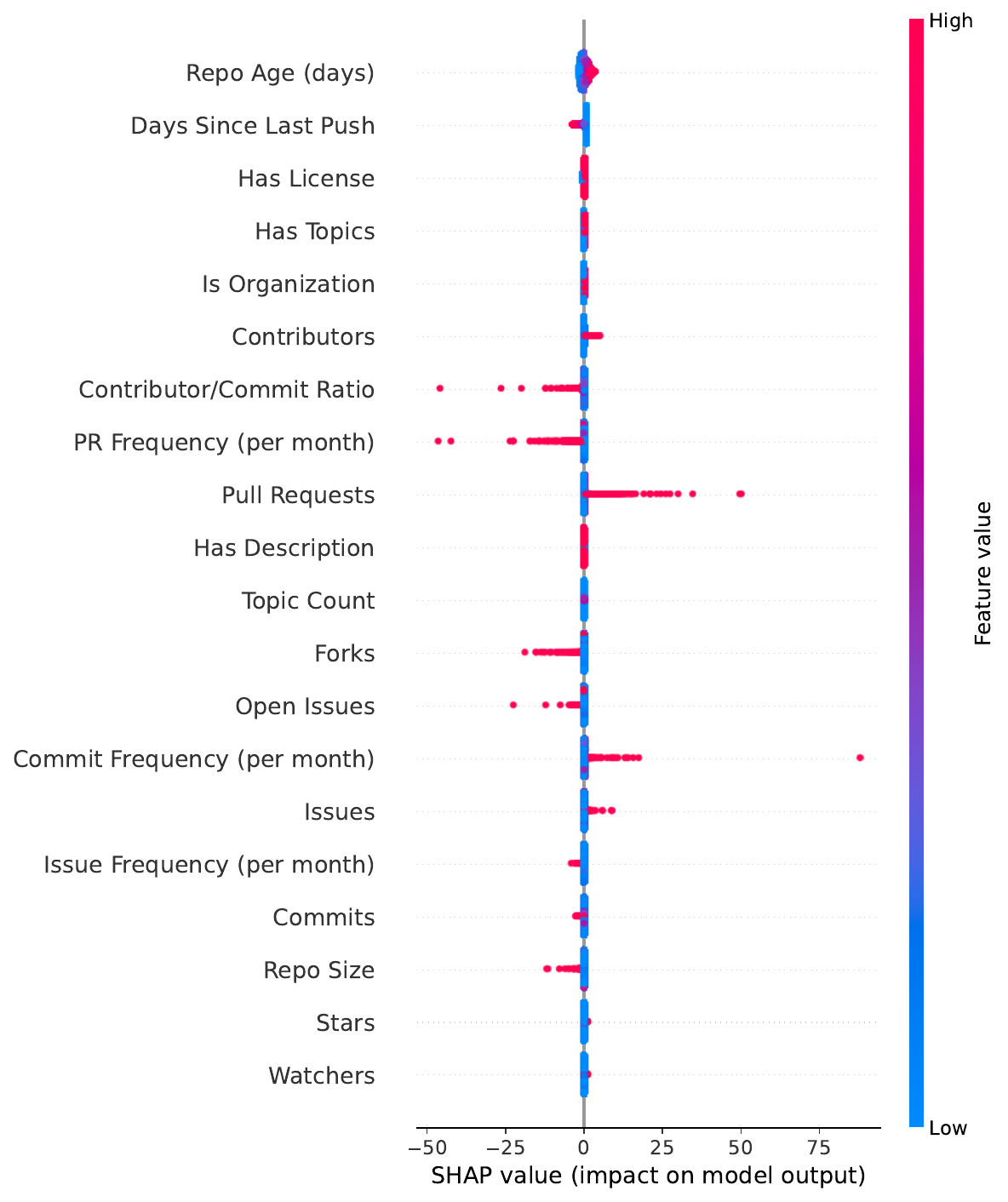}
\Description{SHAP summary plot showing the contribution of repository-level features to the multi-CI adoption model. Features are ordered by mean absolute SHAP value. Each point represents a repository, with horizontal position indicating the feature's contribution to the predicted adoption probability and color indicating the feature value from low to high. Repository age and days since last push have the largest effects, with older repositories and more recent activity generally associated with higher adoption likelihood. Pull-request frequency and contributor-to-commit ratio show effects opposite to raw pull-request count.}
\caption{SHAP summary plot for the multi-CI adoption model, with features ordered by mean absolute SHAP value.}
\label{fig:rq4_shap_adoption}
\end{figure}

The logistic model achieves a pseudo-$R^2$ of 0.188, meaningful for a sociotechnical outcome (Table~\ref{tab:model1}). Repository age is the dominant predictor (OR 2.45, SHAP 34.3\%), with days since last push second and negative (OR 0.43, SHAP 27.6\%). Adoption is thus a behavior of active, long-lived projects, and age and recency together account for over 60\% of the model's explanatory weight. Workflow features add nuance: a high absolute volume of pull requests predicts adoption (OR 1.31), but a high PR rate relative to project lifespan predicts against it (OR 0.76), suggesting that high-velocity teams standardize on one system to avoid pipeline conflicts. Organizational ownership (OR 1.16), a license (OR 1.27), and topics (OR 1.21) all raise the odds, indicating that formally governed projects diversify more, while forks (OR 0.85) and open issues (OR 0.93) lower them. The per-repository contributions in Fig.~\ref{fig:rq4_shap_adoption} make the direction of these effects visible, with the color gradient on repository age and days since last push separating the older, more recently active adopters from the rest.

\begin{table}[htbp]
\centering
\caption{Model 1, multi-CI adoption at the repo level (pseudo-$R^2 = 0.188$). Top predictors by SHAP importance.}
\label{tab:model1}
\begin{tabular}{p{4cm}rrr}
\toprule
\textbf{Feature} & \textbf{OR} & \textbf{95\% CI} & \textbf{SHAP \%} \\
\midrule
Repository age          & 2.45 & [2.41, 2.50] & 34.3 \\
Days since last push    & 0.43 & [0.42, 0.44] & 27.6 \\
Pull requests           & 1.31 & --           & --   \\
Contributor/commit ratio & 0.77 & --          & --   \\
PR frequency (per month) & 0.76 & --          & --   \\
Has license             & 1.27 & --           & --   \\
Has topics              & 1.21 & --           & --   \\
Organization owner      & 1.16 & --           & --   \\
Forks                   & 0.85 & --           & --   \\
Open issues             & 0.93 & --           & --   \\
\bottomrule
\end{tabular}
\end{table}

\subsubsection{RQ4.2: What predicts CI abandonment}
\begin{figure}[htbp]
\centering
\includegraphics[width=0.7\linewidth]{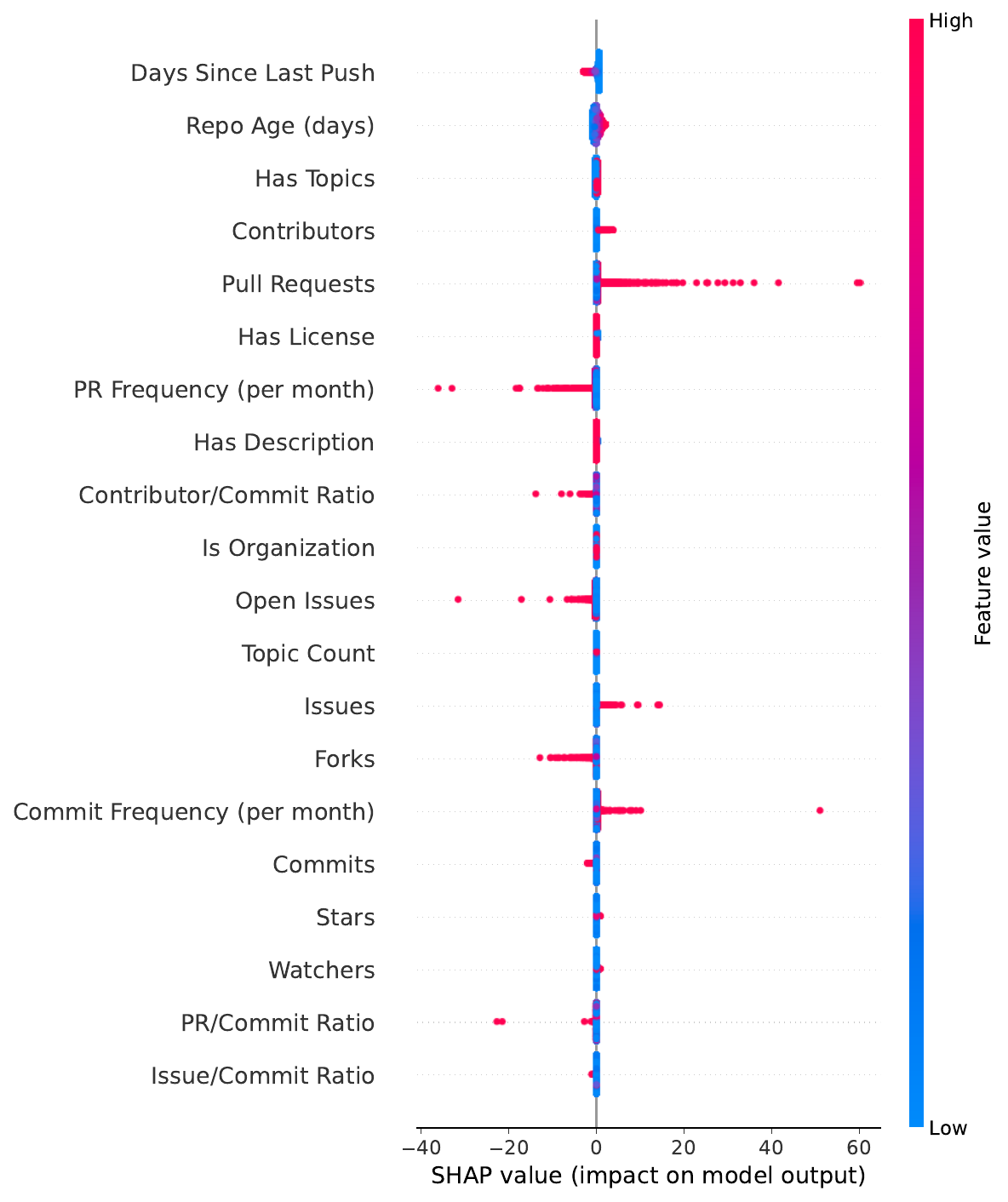}
\caption{SHAP summary plot for the repository-level CI abandonment model, with features ordered by mean absolute SHAP value.}
\Description{Horizontal SHAP summary plot showing the distribution of feature contributions for the repository-level CI abandonment model. Features are ordered by mean absolute SHAP value. Days since last push and repository age have the largest variation and the greatest contribution to model predictions. Higher values of pull-request count are generally associated with positive contributions toward predicted abandonment, while higher pull-request frequency is generally associated with negative contributions. Other repository characteristics have smaller and more concentrated effects.}
\label{fig:rq4_shap_abandon}
\end{figure}

At the repo level, abandoned repositories are more active than retained ones (median commits 208 vs. 81, $r = 0.32$), since active repositories were more likely to adopt CI and so have more to abandon. The model achieves a pseudo-$R^2$ of 0.107, again dominated by repository age (OR 1.82) and days since last push (OR 0.50), together 67\% of SHAP importance. These two findings should be interpreted cautiously, since abandonment is defined partly through the absence of recent CI configuration activity. Older and less recently pushed repositories may therefore be labeled as abandoned partly by construction, and the remaining predictors should be interpreted as stronger indicators of abandonment patterns beyond this definitional effect. Among those, having a license (OR 0.97) and a description (OR 0.96) reduce abandonment odds, consistent with intentional stewardship, while pull requests remain a positive predictor and PR frequency a negative one, mirroring RQ4.1 (Fig.~\ref{fig:rq4_shap_abandon}).

The repo-service model has the lowest fit (pseudo-$R^2 = 0.054$), which is expected given that service-level abandonment is driven by platform events outside any repository's features. The service indicators, rather than any repository feature, are its most consequential coefficients (Table~\ref{tab:model3}). Their magnitudes are sensitive to how the modeling subset is sampled. We therefore read them for structure here and establish their direction from the full-population refit and the Cox model of RQ2.5, as RQ4.3 details. Travis CI is the strongest predictor of service abandonment (OR 1.56, SHAP 22.6\%), a direct imprint of its 2020--2022 removal of free open-source access. CircleCI is second (OR 1.43, SHAP 12.0\%). Together, they account for over a third of total importance, establishing that platform policy dominates repository behavior in explaining which services are dropped. GitHub Actions also has a positive coefficient (OR 1.30), reflecting its role as the destination of migrations rather than a tendency to be abandoned. A repository that migrated from Travis CI to GitHub Actions contributes one abandoned Travis CI record and one retained GitHub Actions record. GitHub Actions adopters are also disproportionately projects in active transition. GitLab CI instead lowers abandonment odds (OR 0.95), consistent with its tight coupling to the GitLab platform.

\begin{table}[t]
\centering
\caption{Model 3, service-level abandonment (pseudo-$R^2 = 0.054$). Service-identity predictors by SHAP importance.}
\label{tab:model3}
\begin{tabular}{p{4cm}rrr}
\toprule
\textbf{Feature} & \textbf{OR} & \textbf{95\% CI} & \textbf{SHAP \%} \\
\midrule
Travis CI (service)     & 1.56 & [1.50, 1.62] & 22.6 \\
CircleCI (service)      & 1.43 & [1.38, 1.48] & 12.0 \\
GitHub Actions (service) & 1.30 & [1.25, 1.35] & 9.9 \\
Azure Pipelines (service) & 1.14 & --         & --  \\
GitLab CI (service)     & 0.95 & --           & --  \\
Days since last push    & 0.76 & --           & --  \\
Repository age          & 0.83 & --           & --  \\
Has license             & 0.87 & --           & --  \\
\bottomrule
\end{tabular}
\end{table}

\subsubsection{RQ4.3: Discrimination and robustness}
Out-of-sample discrimination confirms that these models capture real signal rather than in-sample overfit. Under 5-fold cross-validation, the multi-CI adoption model achieves a ROC-AUC of 0.800 ($\pm 0.003$), a PR-AUC of 0.497 vs. a 0.19 base rate, and a Brier score of 0.179. It is therefore both discriminating and reasonably calibrated. The repo-level abandonment model is weaker but still informative (ROC-AUC 0.718, PR-AUC 0.433 vs. a 0.20 base rate). This result is consistent with abandonment being driven partly by platform events that no repository feature can capture. The variance inflation factors are all modest after the correlation pruning (maximum 5.69, for the issue count, with none approaching the threshold of 10). The odds ratios are not destabilized by collinearity. Replacing logistic regression with a gradient-boosting classifier raises the AUC only modestly: from 0.800 to 0.843 for adoption and from 0.718 to 0.776 for abandonment. The linear model therefore already captures most of the structure, and the reported odds ratios are not masking strong nonlinear effects.

The repo-service abandonment model requires the most caution, and the sensitivity refit is the reason we foreground the service dummies but not their magnitudes. Refitting on the full population of \num{165039} pairs (18.9\% abandonment) rather than the enriched subset (79.6\%) preserves the sign of only 15 of the 27 shared coefficients (56\% agreement). The consequential changes are concentrated in the service indicators that drive the model's primary effect. Travis CI and CircleCI, the two strongest positive predictors in the enriched subset, collapse to near zero. GitHub Actions swings from a small positive coefficient to a strongly negative one and becomes the largest-magnitude service effect in the full population. The remaining sign changes fall on coefficients that are near zero in both fits, where the sign provides no interpretation. The stable conclusion is structural rather than numerical. In both fits, a service indicator, rather than any repository feature, is the largest-magnitude predictor of service-level abandonment. However, which service and in which direction depends on the sampling. We therefore treat the full-population fit as authoritative for direction. The elevated Travis CI and CircleCI hazards in the Cox model of RQ2.5 (HR 1.52 and 1.66) corroborate that these services exhibit higher abandonment hazard beyond the enriched logistic fit. That model is fit on the full \num{161309}-pair population with survival adjustment. We base our interpretation on the survival model rather than the enriched-subset coefficients. We discuss the selection issue further in Section~\ref{sec:threats}.

\begin{rqsummary}{RQ4 Summary}
Multi-CI adoption is best explained by project maturity, as repository age and recent activity together account for most of the predictive signal, which makes age a potential confounding factor that studies of CI diversity should control for. Which service a project abandons is associated with the service's own identity, with Travis CI and CircleCI dwarfing any repository characteristic. CI retention is as much a function of vendor behavior as of project needs.
\end{rqsummary}

\section{Discussion}
\label{sec:discussion}

\noindent\textbf{Service, not language, is the axis of CI practice.}
The most consistent result across RQ1 and RQ2 is that the CI service, rather than the programming language, determines how CI is adopted, configured, maintained, and abandoned. Across adoption counts, maintenance effort, configuration complexity, and developer engagement, language effects are negligible to small once effect sizes are considered, while service-level differences are large. The exceptions are interpretable and narrow, above all the concentration of AppVeyor in the Windows-oriented C, C++, and Rust languages and the language-sensitive verbosity of Travis CI. These findings suggest that broadening language coverage remains useful for validating the generality of CI practices. However, future studies should account for the dominant role of service choice when interpreting cross-language differences. For practitioners and tool builders, this means that guidance and tooling for CI selection and migration can be organized around services rather than fragmented by language. Researchers comparing CI practices across languages should expect the service mix, not the language, to explain most of what they observe.

\noindent\textbf{Multi-CI adoption is largely transitional and often leaves configuration debt.}
Although about one in five repositories adopt multiple services over their lifetime, far fewer keep several simultaneously active, and the concurrent combinations are dominated by GitHub Actions paired with a legacy service. Much multi-CI use is therefore the visible middle of a migration rather than a deliberate steady state, a reading reinforced by RQ3, where co-adoption rationale is almost absent from commit messages. The residue of these transitions is a real cost: abandoned and obsolete configurations accumulate, with AppVeyor reaching a combined lifecycle loss near 68\%, and post-migration cleanup often lags the decision that prompted it. Repositories are left with stale CI files that clutter history and can mislead contributors. Better support for retiring CI configurations, and for detecting stale ones, would address a debt that our data shows is widespread. Automating the migration step itself is an emerging complementary direction, with recent work proposing to translate CI configurations between services using large language models~\cite{cigrate}.

\noindent\textbf{Platform policy changes can reshape CI adoption and abandonment.}
The single largest signal in the lifecycle data is the collapse of Travis CI following its 2020 move to a paid model, which is the dominant abandonment event in every language (RQ2.4), shapes the steepest survival declines (RQ2.5), and is the strongest single predictor of service-level abandonment in the model (RQ4). A commercial decision by one provider reshaped CI practice across the entire open-source population we study, more than any repository characteristic. The flip side is consolidation around GitHub Actions, whose durability and near-universal role as a migration destination raise the familiar concern of platform dependence, now concentrated in a single vendor tightly coupled to the hosting platform.

\noindent\textbf{CI work is mostly reactive, and its costs are under-recorded.}
RQ3 shows that CI configuration activity is mainly maintenance rather than strategy, with reactive deprecation and reliability work alone accounting for more than a third of explanatory commits. This reframes CI maintenance as a continuous operational burden punctuated by occasional strategic decisions, rather than a series of deliberate choices. It also exposes a blind spot for repository mining: developers record symptoms such as build minutes and memory limits but never the pricing behind them. Cost-driven CI decisions, which RQ4 shows are decisive at the platform level, are systematically invisible to commit-message analysis and must be explored through other sources.

\noindent\textbf{Commit messages reveal maintenance work more clearly than strategy.}
A motivation for RQ3 was to test how far the reasons behind CI decisions can be uncovered directly from open-source artifacts, at a scale that interview studies cannot reach. Our results reveal a clear distinction between operational and strategic rationales. Operational reasons, namely deprecations, breakage, and resource limits, are stated readily and extracted reliably, and our three methods converge on them. Strategic reasons, above all why a project co-adopts or chooses a particular destination service, are largely absent from commit messages, since they are typically settled in design discussions, pull-request threads, and issue debates before being enacted through a concise configuration commit. This explains why earlier work studied these motivations through developer interviews~\cite{rostami2023usage}, and our results indicate that interviews and surveys of open-source maintainers remain the appropriate instruments when the object of study is the decision itself. Mining commit messages can quantify the operational evolution of CI at scale, but it cannot replace direct engagement with developers when investigating the reasoning behind strategic choices. A productive division of labor therefore combines large-scale artifact mining, which measures what happens and how often, with targeted surveys or interviews, which explain why.

\section{Threats to Validity}
\label{sec:threats}

\subsection{Construct Validity}
We detect CI adoption from configuration files, which captures configured rather than necessarily executed services, since a repository may retain an unused file or run a service that we cannot infer from repository contents. We proxy configuration complexity by YAML line count, which does not capture semantic or logical complexity, and maintenance effort by the ratio of CI-configuration commits to total commits, which treats all commits as equal. Abandonment is defined by the absence of a configuration file in the current snapshot rather than by an explicit removal event. RQ2.4 uses the last CI commit as a proxy for the removal date, and RQ3 shows that post-migration cleanup can place that date well after the actual decision. For RQ3, commit messages express intent incompletely, which inflates the no-detectable-reason category, and the observed strategic-decision rate within the explanatory subset thus understates the true share and is a lower bound for that subset. Read instead as an estimate for all CI activity, the same rate is an upper bound, since the two-condition filter retains only 1.9\% of CI commits and those reason-bearing commits over-represent strategic decisions relative to the full dataset. The primary labels in RQ3 come from manual coding by the first co-author, which is subject to the subjectivity inherent in qualitative classification. We mitigate this through an explicit taxonomy, independent re-labeling of a stratified 381-commit subset by a second author that reproduced the strategic-versus-maintenance split with 97.4\% agreement and reached almost-perfect agreement on the full taxonomy ($\kappa = 0.85$), and convergent LLM-consensus and topic-modeling analyses that reproduce the central distribution. Residual disagreement concentrates on whether a commit records a specific reason or only a generic configuration edit, the boundary that inherently limits any reason-discovery taxonomy.

\subsection{Internal Validity}
Adoption and switching timelines are reconstructed from commit history, which squashing, rebasing, or repository restructuring can distort. Developer counts rest on GitHub logins, which alias inconsistently across a contributor's identities. In the RQ4 abandonment models, repository age and recency of activity are partly entangled with the abandonment definition itself. We disclose this circularity and treat the remaining predictors as more informative of factors associated with abandonment beyond repository inactivity. The repo-service modeling subset is enriched in abandonment (79.6\%) relative to the full population (18.9\%), as it retains only pairs with sufficient history and features, which is why its coefficients describe which services are dropped among selected pairs rather than population abandonment rates. Refitting that model on the full population confirms the concern is real: only 56\% of coefficient signs are preserved, and the large-magnitude changes fall on the service indicators. The enriched subset's absolute service effects do not transfer, even though a service indicator remains the largest-magnitude predictor under both fits. We therefore rely on the full-population fit for the direction of service effects, treat the enriched-subset magnitudes as descriptive of that subset only, and base the service-level abandonment ranking on the Cox model of RQ2.5, which is fit on the full pair population with survival adjustment.

\subsection{External Validity}
Our study covers public GitHub repositories in seven languages that adopted one of eight CI services. It does not capture self-hosted or private CI, services outside our set, or repositories hosted elsewhere such as on GitLab, where adoption dynamics may differ. The seven languages were chosen to span compiled and interpreted ones, but other languages and application domains may behave differently. The reasons studied in RQ3 come from commit messages, and richer rationale, especially for co-adoption, resides in pull-request and issue discussions we did not mine. Our findings should be applied with these boundaries in mind.

\subsection{Conclusion Validity}
Our metrics are heavy-tailed, which is why we use nonparametric tests throughout, following established guidance for statistical testing in software engineering~\cite{arcuri2011}, and report the median as the primary statistic alongside means and upper percentiles. Given that the large sample size makes trivial differences statistically significant, we pair every test with an effect size and interpret practical magnitude rather than $p$-values, which is why several statistically significant language effects are reported as negligible. We apply Holm, Benjamini--Hochberg, or Bonferroni correction wherever we run families of tests. Some services, namely Bitbucket Pipelines, Cirrus CI, and Azure Pipelines, have small per-language samples, which leaves their cross-language comparisons underpowered, and we flag them as such rather than drawing strong conclusions. The LLM consensus and BERTopic analyses in RQ3 are treated as directional and exploratory validation, not as confirmatory evidence. In the RQ2.5 Cox model, the scaled Schoenfeld residual test rejects the proportional-hazards assumption for most covariates, but with over \num{161000} observations the test flags even minor, practically irrelevant deviations, as its authors caution~\cite{grambsch1994}. We therefore read the hazard ratios as average effects over the observation window rather than as constant instantaneous effects, and we report restricted mean survival time, which needs no proportional-hazards assumption, as the primary retention summary alongside them.

\section{Conclusion}
\label{sec:conclusion}
We presented a large-scale longitudinal study of multi-CI service adoption across seven programming languages and eight CI services, spanning \num{135227} GitHub repositories. Our results show that CI service choice, rather than programming language, is the dominant factor shaping adoption, configuration complexity, maintenance, and evolution. Most repositories rely on a single service, while GitHub Actions has consolidated first adoptions across all studied languages and become the central platform in the current CI landscape. Language effects are limited to specific cases, such as the concentration of AppVeyor among Windows-oriented C, C++, and Rust projects.
We further showed that CI evolution is strongly influenced by service lifecycle events. About 19\% of repositories switch CI services over time, with most migrations moving toward GitHub Actions. The decline of Travis CI following its 2020 pricing change represents the strongest abandonment signal in our data and demonstrates how vendor decisions can reshape repository practices. Multi-CI adoption is primarily associated with project maturity, whereas abandonment is driven more by service identity than by repository characteristics.
By examining the rationale recorded in CI configuration commits, we found that CI activity is mostly reactive: only about 11\% of explanatory commits are related to strategic decisions, while maintenance responses to deprecations and reliability issues dominate. This highlights both the value and the limitations of commit messages as evidence of developer intent: they reveal how CI configurations evolve, but only partially capture why teams make broader tooling decisions.
Overall, this study shows that CI practice is shaped by the interaction between project maturity, service characteristics, and external platform changes.

\smallskip\noindent\textbf{Future work.}
Future work should explore the strategic rationale behind CI decisions by mining richer sources of developer communication, such as pull-request discussions, issue threads, and code review conversations, which can capture motivations that are not always reflected in commit messages. These sources are the most promising route to the cost-driven and destination-choice rationale that RQ3 finds absent from commit messages, building on evidence that latent design rationale can be located in pull-request discussions~\cite{viviani2021}. Extending the analysis to additional hosting platforms, such as GitLab and Bitbucket, and to self-hosted CI deployments, such as Jenkins and TeamCity installations managed within organizations, would help determine whether the observed patterns generalize beyond GitHub-hosted projects. Another important direction is to connect CI configuration evolution with engineering outcomes, including build reliability, failure rates, execution cost, and developer productivity, to better understand the consequences of different CI practices. Finally, the predictors identified in this study could serve as a foundation for recommendation systems that assist developers in selecting, adopting, migrating, or consolidating CI services based on repository characteristics and project needs.

\begin{acks}
This work is funded by the Natural Sciences and Engineering Research Council of Canada (NSERC): RGPIN-2025-05897. The study was partly enabled by the Digital Research Alliance of Canada.
\end{acks}

\section*{Data Availability}
Data, analysis scripts, and full results that support the findings of this study are available in our replication package~\cite{our_replication_package}.

\bibliographystyle{ACM-Reference-Format}
\bibliography{paper}
\end{document}